\documentclass[aps,prd,twocolumn,showpacs,floatfix,superscriptaddress]{revtex4}

\usepackage{amsmath}
\usepackage{amssymb}
\usepackage{color}
\usepackage{dcolumn}
\usepackage{graphicx}
\usepackage{multirow}
\usepackage{rotating}

\newcolumntype{.}[1]{D{.}{.}{#1}}

\definecolor{light-gray}{gray}{0.6}

\def\mathbi#1{\textbf{\em #1}}

\newcommand{\dzero}  {D0}
\newcommand{\met}    {\mbox{$\not\!\!E_T$}}
\newcommand{\rar}    {\rightarrow}
\newcommand{\rargap} {\mbox{ $\rightarrow$ }}

\newcommand{\nap}    {\multicolumn{1}{c}{n/a}}
\newcommand{\nev}    {\multicolumn{1}{c}{n/e}}
\newcommand{\zp}     {0}

\newcommand{\ttbar}  {\ensuremath{t\bar{t}}}

\newcommand{\z}      {0.00}
\newcommand{\jk}     {0.24}
\newcommand{\jj}     {0.26}

\newcommand{\tr}     {0.06}
\newcommand{\bb}     {0.10}

\newcommand{\pd}     {0.14}
\newcommand{\gf}     {0.03}
\newcommand{\is}     {0.15}
\newcommand{\jh}     {0.25}
\newcommand{\bu}     {0.16}
\newcommand{\bm}     {0.16}
\newcommand{\cs}     {0.37}
\newcommand{\pe}     {0.24}
\newcommand{\iu}     {0.26}
\newcommand{\ho}     {0.25}
\newcommand{\ji}     {0.58}
\newcommand{\bv}     {0.04}
\newcommand{\bn}     {0.06}
\newcommand{\ct}     {0.28}

\newcommand{\hj}     {0.03}
\newcommand{\qs}     {0.07}
\newcommand{\ti}     {0.25}
\newcommand{\bs}     {0.07}
\newcommand{\hk}     {0.07}
\newcommand{\qt}     {0.16}
\newcommand{\tj}     {0.07}
\newcommand{\bt}     {0.03}

\newcommand{\sg}     {0.10}
\newcommand{\mc}     {0.14}

\begin{document}

\hspace{5.2in} \mbox{FERMILAB-PUB-12-336-E}


%
%
%

\title{\vspace{0.25in}Combination of the top-quark mass measurements from the Tevatron collider\vspace{0.1in}}
\affiliation{LAFEX, Centro Brasileiro de Pesquisas F\'{i}sicas, Rio de Janeiro, Brazil}
\affiliation{Universidade do Estado do Rio de Janeiro, Rio de Janeiro, Brazil}
\affiliation{Universidade Federal do ABC, Santo Andr\'e, Brazil}
\affiliation{Institute of Particle Physics: McGill University, Montr\'{e}al, Qu\'{e}bec, Canada H3A~2T8; Simon Fraser University, Burnaby, British Columbia, Canada V5A~1S6; University of Toronto, Toronto, Ontario, Canada M5S~1A7; and TRIUMF, Vancouver, British Columbia, Canada V6T~2A3}
\affiliation{University of Science and Technology of China, Hefei, People's Republic of China}
\affiliation{Institute of Physics, Academia Sinica, Taipei, Taiwan 11529, Republic of China}
\affiliation{Universidad de los Andes, Bogot\'a, Colombia}
\affiliation{Charles University, Faculty of Mathematics and Physics, Center for Particle Physics, Prague, Czech Republic}
\affiliation{Czech Technical University in Prague, Prague, Czech Republic}
\affiliation{Center for Particle Physics, Institute of Physics, Academy of Sciences of the Czech Republic, Prague, Czech Republic}
\affiliation{Universidad San Francisco de Quito, Quito, Ecuador}
\affiliation{Division of High Energy Physics, Department of Physics, University of Helsinki and Helsinki Institute of Physics, FIN-00014, Helsinki, Finland}
\affiliation{LPC, Universit\'e Blaise Pascal, CNRS/IN2P3, Clermont, France}
\affiliation{LPSC, Universit\'e Joseph Fourier Grenoble 1, CNRS/IN2P3, Institut National Polytechnique de Grenoble, Grenoble, France}
\affiliation{CPPM, Aix-Marseille Universit\'e, CNRS/IN2P3, Marseille, France}
\affiliation{LAL, Universit\'e Paris-Sud, CNRS/IN2P3, Orsay, France}
\affiliation{LPNHE, Universit\'es Paris VI and VII, CNRS/IN2P3, Paris, France}
\affiliation{CEA, Irfu, SPP, Saclay, France}
\affiliation{IPHC, Universit\'e de Strasbourg, CNRS/IN2P3, Strasbourg, France}
\affiliation{IPNL, Universit\'e Lyon 1, CNRS/IN2P3, Villeurbanne, France and Universit\'e de Lyon, Lyon, France}
\affiliation{III. Physikalisches Institut A, RWTH Aachen University, Aachen, Germany}
\affiliation{Physikalisches Institut, Universit\"at Freiburg, Freiburg, Germany}
\affiliation{II. Physikalisches Institut, Georg-August-Universit\"at G\"ottingen, G\"ottingen, Germany}
\affiliation{Institut f\"{u}r Experimentelle Kernphysik, Karlsruhe Institute of Technology, D-76131 Karlsruhe, Germany}
\affiliation{Institut f\"ur Physik, Universit\"at Mainz, Mainz, Germany}
\affiliation{Ludwig-Maximilians-Universit\"at M\"unchen, M\"unchen, Germany}
\affiliation{Fachbereich Physik, Bergische Universit\"at Wuppertal, Wuppertal, Germany}
\affiliation{University of Athens, 157 71 Athens, Greece}
\affiliation{Panjab University, Chandigarh, India}
\affiliation{Delhi University, Delhi, India}
\affiliation{Tata Institute of Fundamental Research, Mumbai, India}
\affiliation{University College Dublin, Dublin, Ireland}
\affiliation{Istituto Nazionale di Fisica Nucleare Bologna, $^{\star{a}}$University of Bologna, I-40127 Bologna, Italy}
\affiliation{Laboratori Nazionali di Frascati, Istituto Nazionale di Fisica Nucleare, I-00044 Frascati, Italy}
\affiliation{Istituto Nazionale di Fisica Nucleare, Sezione di Padova-Trento, $^{\star{b}}$University of Padova, I-35131 Padova, Italy}
\affiliation{Istituto Nazionale di Fisica Nucleare Pisa, $^{\star{c}}$University of Pisa, $^{\star{d}}$University of Siena and $^{\star{e}}$Scuola Normale Superiore, I-56127 Pisa, Italy}
\affiliation{Istituto Nazionale di Fisica Nucleare, Sezione di Roma 1, $^{\star{f}}$Sapienza Universit\`{a} di Roma, I-00185 Roma, Italy}
\affiliation{Istituto Nazionale di Fisica Nucleare Trieste/Udine, I-34100 Trieste, $^{\star{g}}$University of Udine, I-33100 Udine, Italy}
\affiliation{Okayama University, Okayama 700-8530, Japan}
\affiliation{Osaka City University, Osaka 588, Japan}
\affiliation{Waseda University, Tokyo 169, Japan}
\affiliation{University of Tsukuba, Tsukuba, Ibaraki 305, Japan}
\affiliation{Center for High Energy Physics: Kyungpook National University, Daegu 702-701, Korea; Seoul National University, Seoul 151-742, Korea; Sungkyunkwan University, Suwon 440-746, Korea; Korea Institute of Science and Technology Information, Daejeon 305-806, Korea; Chonnam National University, Gwangju 500-757, Korea; Chonbuk National University, Jeonju 561-756, Korea}
\affiliation{Korea Detector Laboratory, Korea University, Seoul, Korea}
\affiliation{CINVESTAV, Mexico City, Mexico}
\affiliation{Nikhef, Science Park, Amsterdam, the Netherlands}
\affiliation{Radboud University Nijmegen, Nijmegen, the Netherlands}
\affiliation{Joint Institute for Nuclear Research, Dubna, Russia}
\affiliation{Institute for Theoretical and Experimental Physics, Moscow, Russia}
\affiliation{Moscow State University, Moscow, Russia}
\affiliation{Institute for High Energy Physics, Protvino, Russia}
\affiliation{Petersburg Nuclear Physics Institute, St. Petersburg, Russia}
\affiliation{Comenius University, 842 48 Bratislava, Slovakia; Institute of Experimental Physics, 040 01 Kosice, Slovakia}
\affiliation{Institut de Fisica d'Altes Energies, ICREA, Universitat Autonoma de Barcelona, E-08193, Bellaterra (Barcelona), Spain}
\affiliation{Instituci\'{o} Catalana de Recerca i Estudis Avan\c{c}ats (ICREA) and Institut de F\'{i}sica d'Altes Energies (IFAE), Barcelona, Spain}
\affiliation{Centro de Investigaciones Energeticas Medioambientales y Tecnologicas, E-28040 Madrid, Spain}
\affiliation{Instituto de Fisica de Cantabria, CSIC-University of Cantabria, 39005 Santander, Spain}
\affiliation{Uppsala University, Uppsala, Sweden}
\affiliation{University of Geneva, CH-1211 Geneva 4, Switzerland}
\affiliation{Glasgow University, Glasgow G12 8QQ, United Kingdom}
\affiliation{Lancaster University, Lancaster LA1 4YB, United Kingdom}
\affiliation{University of Liverpool, Liverpool L69 7ZE, United Kingdom}
\affiliation{Imperial College London, London SW7 2AZ, United Kingdom}
\affiliation{University College London, London WC1E 6BT, United Kingdom}
\affiliation{The University of Manchester, Manchester M13 9PL, United Kingdom}
\affiliation{University of Oxford, Oxford OX1 3RH, United Kingdom}
\affiliation{University of Arizona, Tucson, Arizona 85721, USA}
\affiliation{Ernest Orlando Lawrence Berkeley National Laboratory, Berkeley, California 94720, USA}
\affiliation{University of California, Davis, Davis, California 95616, USA}
\affiliation{University of California, Los Angeles, Los Angeles, California 90024, USA}
\affiliation{University of California Riverside, Riverside, California 92521, USA}
\affiliation{Yale University, New Haven, Connecticut 06520, USA}
\affiliation{University of Florida, Gainesville, Florida 32611, USA}
\affiliation{Florida State University, Tallahassee, Florida 32306, USA}
\affiliation{Argonne National Laboratory, Argonne, Illinois 60439, USA}
\affiliation{Fermi National Accelerator Laboratory, Batavia, Illinois 60510, USA}
\affiliation{Enrico Fermi Institute, University of Chicago, Chicago, Illinois 60637, USA}
\affiliation{University of Illinois at Chicago, Chicago, Illinois 60607, USA}
\affiliation{Northern Illinois University, DeKalb, Illinois 60115, USA}
\affiliation{Northwestern University, Evanston, Illinois 60208, USA}
\affiliation{University of Illinois, Urbana, Illinois 61801, USA}
\affiliation{Indiana University, Bloomington, Indiana 47405, USA}
\affiliation{Purdue University Calumet, Hammond, Indiana 46323, USA}
\affiliation{University of Notre Dame, Notre Dame, Indiana 46556, USA}
\affiliation{Purdue University, West Lafayette, Indiana 47907, USA}
\affiliation{Iowa State University, Ames, Iowa 50011, USA}
\affiliation{University of Kansas, Lawrence, Kansas 66045, USA}
\affiliation{Kansas State University, Manhattan, Kansas 66506, USA}
\affiliation{Louisiana Tech University, Ruston, Louisiana 71272, USA}
\affiliation{The Johns Hopkins University, Baltimore, Maryland 21218, USA}
\affiliation{Boston University, Boston, Massachusetts 02215, USA}
\affiliation{Northeastern University, Boston, Massachusetts 02115, USA}
\affiliation{Harvard University, Cambridge, Massachusetts 02138, USA}
\affiliation{Massachusetts Institute of Technology, Cambridge, Massachusetts 02139, USA}
\affiliation{Tufts University, Medford, Massachusetts 02155, USA}
\affiliation{University of Michigan, Ann Arbor, Michigan 48109, USA}
\affiliation{Wayne State University, Detroit, Michigan 48201, USA}
\affiliation{Michigan State University, East Lansing, Michigan 48824, USA}
\affiliation{University of Mississippi, University, Mississippi 38677, USA}
\affiliation{University of Nebraska, Lincoln, Nebraska 68588, USA}
\affiliation{Rutgers University, Piscataway, New Jersey 08855, USA}
\affiliation{Princeton University, Princeton, New Jersey 08544, USA}
\affiliation{University of New Mexico, Albuquerque, New Mexico 87131, USA}
\affiliation{State University of New York, Buffalo, New York 14260, USA}
\affiliation{The Rockefeller University, New York, New York 10065, USA}
\affiliation{University of Rochester, Rochester, New York 14627, USA}
\affiliation{State University of New York, Stony Brook, New York 11794, USA}
\affiliation{Brookhaven National Laboratory, Upton, New York 11973, USA}
\affiliation{Duke University, Durham, North Carolina 27708, USA}
\affiliation{The Ohio State University, Columbus, Ohio 43210, USA}
\affiliation{Langston University, Langston, Oklahoma 73050, USA}
\affiliation{University of Oklahoma, Norman, Oklahoma 73019, USA}
\affiliation{Oklahoma State University, Stillwater, Oklahoma 74078, USA}
\affiliation{University of Pennsylvania, Philadelphia, Pennsylvania 19104, USA}
\affiliation{Carnegie Mellon University, Pittsburgh, Pennsylvania 15213, USA}
\affiliation{University of Pittsburgh, Pittsburgh, Pennsylvania 15260, USA}
\affiliation{Brown University, Providence, Rhode Island 02912, USA}
\affiliation{University of Texas, Arlington, Texas 76019, USA}
\affiliation{Texas A\&M University, College Station, Texas 77843, USA}
\affiliation{Southern Methodist University, Dallas, Texas 75275, USA}
\affiliation{Rice University, Houston, Texas 77005, USA}
\affiliation{Baylor University, Waco, Texas 76798, USA}
\affiliation{University of Virginia, Charlottesville, Virginia 22904, USA}
\affiliation{University of Washington, Seattle, Washington 98195, USA}
\affiliation{University of Wisconsin, Madison, Wisconsin 53706, USA}
\author{T.~Aaltonen} \affiliation{Division of High Energy Physics, Department of Physics, University of Helsinki and Helsinki Institute of Physics, FIN-00014, Helsinki, Finland}
\author{V.M.~Abazov} \affiliation{Joint Institute for Nuclear Research, Dubna, Russia}
\author{B.~Abbott} \affiliation{University of Oklahoma, Norman, Oklahoma 73019, USA}
\author{B.S.~Acharya} \affiliation{Tata Institute of Fundamental Research, Mumbai, India}
\author{M.~Adams} \affiliation{University of Illinois at Chicago, Chicago, Illinois 60607, USA}
\author{T.~Adams} \affiliation{Florida State University, Tallahassee, Florida 32306, USA}
\author{G.D.~Alexeev} \affiliation{Joint Institute for Nuclear Research, Dubna, Russia}
\author{G.~Alkhazov} \affiliation{Petersburg Nuclear Physics Institute, St. Petersburg, Russia}
\author{A.~Alton$^{{\ddag}a}$} \affiliation{University of Michigan, Ann Arbor, Michigan 48109, USA}
\author{B.~\'{A}lvarez~Gonz\'{a}lez$^{{\dag}z}$} \affiliation{Instituto de Fisica de Cantabria, CSIC-University of Cantabria, 39005 Santander, Spain}
% ...
~\affiliation{Fermi National Accelerator Laboratory, Batavia, Illinois 60510, USA}
\author{F.~Rizatdinova} \affiliation{Oklahoma State University, Stillwater, Oklahoma 74078, USA}
\author{A.~Robson} \affiliation{Glasgow University, Glasgow G12 8QQ, United Kingdom}
\author{T.~Rodrigo} \affiliation{Instituto de Fisica de Cantabria, CSIC-University of Cantabria, 39005 Santander, Spain}
\author{T.~Rodriguez} \affiliation{University of Pennsylvania, Philadelphia, Pennsylvania 19104, USA}
\author{E.~Rogers} \affiliation{University of Illinois, Urbana, Illinois 61801, USA}
\author{S.~Rolli$^{{\dag}i}$} \affiliation{Tufts University, Medford, Massachusetts 02155, USA}
\author{M.~Rominsky} \affiliation{Fermi National Accelerator Laboratory, Batavia, Illinois 60510, USA}
\author{R.~Roser} \affiliation{Fermi National Accelerator Laboratory, Batavia, Illinois 60510, USA}
\author{A.~Ross} \affiliation{Lancaster University, Lancaster LA1 4YB, United Kingdom}
\author{C.~Royon} \affiliation{CEA, Irfu, SPP, Saclay, France}
% ...

\collaboration{CDF and D0 Collaborations
\footnote{With CDF visitors from
$^{{\dag}a}$Istituto Nazionale di Fisica Nucleare, Sezione di Cagliari, 09042 Monserrato (Cagliari), Italy,
$^{{\dag}b}$University of California Irvine, Irvine, CA 92697, USA,
$^{{\dag}c}$University of California Santa Barbara, Santa Barbara, CA 93106, USA,
$^{{\dag}d}$University of California Santa Cruz, Santa Cruz, CA 95064, USA,
$^{{\dag}e}$Institute of Physics, Academy of Sciences of the Czech Republic, Czech Republic,
$^{{\dag}f}$CERN, CH-1211 Geneva, Switzerland,
$^{{\dag}g}$Cornell University, Ithaca, NY 14853, USA,
$^{{\dag}h}$University of Cyprus, Nicosia CY-1678, Cyprus,
$^{{\dag}i}$Office of Science, U.S. Department of Energy, Washington, DC 20585, USA,
$^{{\dag}j}$University College Dublin, Dublin 4, Ireland,
$^{{\dag}k}$ETH, 8092 Zurich, Switzerland,
$^{{\dag}l}$University of Fukui, Fukui City, Fukui Prefecture, Japan 910-0017,
$^{{\dag}m}$Universidad Iberoamericana, Mexico D.F., Mexico,
$^{{\dag}n}$University of Iowa, Iowa City, IA 52242, USA,
$^{{\dag}o}$Kinki University, Higashi-Osaka City, Japan 577-8502,
$^{{\dag}p}$Kansas State University, Manhattan, KS 66506, USA,
$^{{\dag}q}$Ewha Womans University, Seoul, 120-750, Korea,
$^{{\dag}r}$University of Manchester, Manchester M13 9PL, United Kingdom,
$^{{\dag}s}$Queen Mary, University of London, London, E1 4NS, United Kingdom,
$^{{\dag}t}$University of Melbourne, Victoria 3010, Australia,
$^{{\dag}u}$Muons, Inc., Batavia, IL 60510, USA,
$^{{\dag}v}$Nagasaki Institute of Applied Science, Nagasaki, Japan,
$^{{\dag}w}$National Research Nuclear University, Moscow, Russia,
$^{{\dag}x}$Northwestern University, Evanston, IL 60208, USA,
$^{{\dag}y}$University of Notre Dame, Notre Dame, IN 46556, USA,
$^{{\dag}z}$Universidad de Oviedo, E-33007 Oviedo, Spain,
$^{{\dag}aa}$CNRS-IN2P3, Paris, F-75205 France,
$^{{\dag}bb}$Texas Tech University, Lubbock, TX 79609, USA,
$^{{\dag}cc}$Universidad Tecnica Federico Santa Maria, 110v Valparaiso, Chile,
$^{{\dag}dd}$Yarmouk University, Irbid 211-63, Jordan,
}
\footnote{and D0 visitors from
$^{{\ddag}a}$Augustana College, Sioux Falls, SD, USA,
$^{{\ddag}b}$The University of Liverpool, Liverpool, UK,
$^{{\ddag}c}$UPIITA-IPN, Mexico City, Mexico,
$^{{\ddag}d}$DESY, Hamburg, Germany,
,
$^{{\ddag}e}$SLAC, Menlo Park, CA, USA,
$^{{\ddag}f}$University College London, London, UK,
$^{{\ddag}g}$Centro de Investigacion en Computacion - IPN, Mexico City, Mexico,
$^{{\ddag}h}$ECFM, Universidad Autonoma de Sinaloa, Culiac\'an, Mexico
and
$^{{\ddag}i}$Universidade Estadual Paulista, S\~ao Paulo, Brazil.
}}
\noaffiliation

\begin{abstract}

The top quark is the heaviest known elementary particle, with a mass about 40 times 
larger than the mass of its isospin partner, the bottom quark. It decays almost 100\% 
of the time to a $W$ boson and a bottom quark.
Using top-antitop pairs at the Tevatron proton-antiproton collider, the
CDF and {\dzero} Collaborations have measured the top
quark's mass in different final states for integrated luminosities of up to 5.8~fb$^{-1}$.
This paper reports on a combination of these measurements
that results in a more precise value of the mass than any
individual decay channel can provide. 
It describes the treatment of the systematic uncertainties and their correlations.
The mass value determined is
$173.18 \pm 0.56 \thinspace ({\rm stat}) \pm 0.75 \thinspace ({\rm syst})$~GeV or  $173.18 \pm
0.94$~GeV, which has a precision of $\pm 0.54\%$, 
making this the most precise determination of the top-quark mass.

\end{abstract}

\pacs{14.65.Ha, 13.85.Ni, 13.85.Qk, 12.15.Ff}

\maketitle


\section{Introduction}
\label{introduction}


\subsection{The top quark}
\label{top-quark}

The standard model (SM) of particle physics describes the elementary
particles and their interactions. 
The top quark ($t$) has a special place in the hierarchy of particles 
because it is far more massive than any of the other fundamental objects. It is the up-type quark, 
partnered with the down-type bottom quark ($b$),  
forming the third generation of quarks that was predicted by Kobayashi and Maskawa in 
1973~\cite{kobayashi-maskawa} to accommodate {\it CP} violation
in neutral kaon decays~\cite{CP-violation}. At particle colliders the top quark is
produced mainly in top-antitop ($t\bar{t}$) pairs. The first evidence of top-quark production was reported
by the CDF Collaboration~\cite{top-evidence-cdf} and the top-quark was first observed in this production  mode  by the
CDF~\cite{top-observation-cdf}, and
{\dzero}~\cite{top-observation-dzero} Collaborations at the Tevatron proton-antiproton
collider. Since then, great efforts have been focused on
measuring its properties with ever higher precision. In addition to
its large mass ($m_t$), the top quark is also singular because it
decays before it can hadronize: there are no mesons or
baryons containing valence top quarks. The top quark decays almost exclusively to a $W$~boson and a
$b$~quark, with the fraction determined by the near-unity value of the 
Cabibbo-Kobayashi-Maskawa (CKM) quark mixing matrix~\cite{kobayashi-maskawa,cabibbo} element
$V_{tb} \left(\approx~0.9992\right)$~\cite{particle-data-book}. Its
other decays are limited by the small values of $V_{ts} \approx 0.0387$
and $V_{td} \approx 0.0084$~\cite{particle-data-book}, assuming
three-family unitarity of the CKM matrix. 
The $W$~boson
decays to a charged lepton and its associated neutrino, or to a
quark-antiquark pair, and the final states of \ttbar\ events are
thus characterized as follows: ``lepton+jets''  ($t \bar{t} \rargap
\ell^+ \nu b q \bar{q}^{\prime} \bar{b}$ and $\bar{q} q^{\prime} b
\ell^- \bar{\nu} \bar{b}$); ``alljets''  ($t \bar{t} \rargap q
\bar{q}^{\prime} b \bar{q} q^{\prime} \bar{b}$), and ``dileptons''
($t \bar{t} \rargap \ell^+ \nu b \ell^- \bar{\nu} \bar{b}$). In
this notation the charged lepton $\ell$ represents an electron or
muon, and $q$ is a first- or second-generation quark. 
The $W$~boson also decays to a $\tau$ lepton and a $\tau$ neutrino. 
If $\tau$ decays to an electron or muon, the event 
contributes to the lepton categories, and if the $\tau$ decays 
into hadrons, it contributes to the lepton+jets or alljets categories.
A fourth category labeled ``{\met}+jets'' is used to measure $m_t$
when there are jets and a large imbalance in transverse momentum in the event ({\met}),
but no identified lepton. It comprises $t \bar{t} \rargap
\tau^+ \nu b \tau^-\bar{\nu} \bar{b}$, $\tau^+ \nu b q
\bar{q}^{\prime} \bar{b}$, and $\bar{q} q^{\prime} b \tau^- \bar{\nu}
\bar{b}$ final states, accounting for 40\% of the \ttbar\ signal events in the {\met}+jets category, 
or $\ell^+ \nu b q \bar{q}^{\prime} \bar{b}$, $\bar{q} q^{\prime} b \ell^- \bar{\nu}
\bar{b}$, where the electron or muon are not reconstructed, accounting for  60\% of the
\ttbar\ signal in this category. 
Additional contributions to {\met} arise from the neutrino(s) produced in $\tau$ decays.

\begin{figure}[!h!tb]
\vspace{-0.1in}
\includegraphics[width=3.2in]{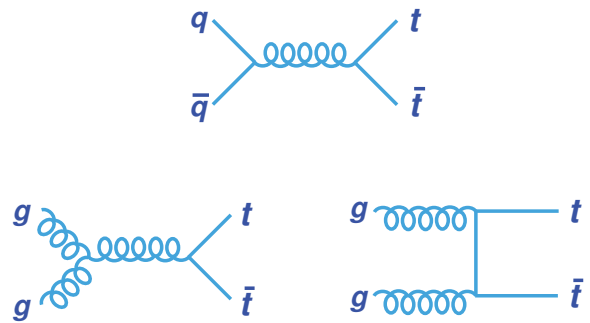}
\vspace{-0.1in}
\caption[feynman1]{Examples of tree Feynman diagrams for $t\bar{t}$
production. At the Tevatron collider, the $q\bar{q}$ channel
contributes 81\% to the total \ttbar\ inclusive cross section and the $gg$ channel the remaining 19\%~\cite{mc@nlo,cteq6}.}
\label{feynman-top-production}
\end{figure}

\begin{figure}[!h!tb]
\vspace{-0.2in}
\includegraphics[width=3.2in]{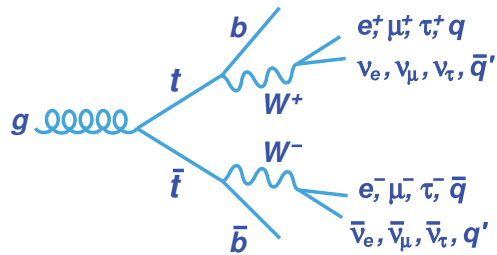}
\vspace{-0.1in}
\caption[feynman2]{Leading-order Feynman diagram for $t\bar{t}$
decay. The dilepton modes ($ee$, $e\mu$, $\mu\mu$) have a combined
branching fraction of $\approx$~4\%, the electron+jets and muon+jets modes combined
correspond to $\approx$~30\%, and the alljets mode has a branching
fraction of $\approx$~46\%. The $\tau$ modes are shared among the {\met}+jets
and the other channels in the analyses.}
\label{feynman-top-decay}
\end{figure}

In dilepton events, there are typically two jets from the two $b$ quarks, one from each top-quark 
decay. In lepton+jets events, there are typically four jets,
including two $b$~jets and two light-quark jets 
from $W$-boson decay. Alljets events
most often contain six jets, the two $b$~jets and four light-quark
jets. The {\met}+jets events usually have four or five jets.
Additional gluon or quark jets can arise owing to radiation from initial or final-state 
colored particles, including the top quarks.
About 23\% of the $t\bar{t}$ events have an extra jet with
sufficient energy to pass the selection criteria,
and about 5\% of the events have two additional
jets. These extra jets complicate the measurement of $m_t$ and
degrade its resolution. Figure~\ref{feynman-top-production}
illustrates leading-order (LO) production of \ttbar\ events at the Fermilab Tevatron Collider,
and Fig.~\ref{feynman-top-decay} shows the relevant \ttbar\ decay modes.

\vspace{-0.2in}
\subsection{Top-quark mass origin and definitions}
\label{nature-mass}

One of the fundamental properties of an elementary particle is its
mass. In the SM, fermions acquire mass through interactions with the 
Higgs field~\cite{higgs-field}. Absolute values of these masses are not
predicted by the SM. 
In theoretical calculations, a particle's mass can be defined in
more than one way, and it depends on how higher-order terms in perturbative quantum
chromodynamics (QCD) calculations are renormalized.
In the modified minimal
subtraction scheme ($\overline{\rm{MS}}$), for example, the mass definition reflects
short-distance effects, whereas in the pole-mass
scheme the mass definition reflects long-distance
effects~\cite{mass-from-xsec-theory}. The concept of the pole mass is
not well defined since color confinement does not provide $S$-matrix
poles at $m = m_t$~\cite{pole-mass-definition}. 
Direct mass measurements that are inputs to the combination described in this paper
rely on Monte Carlo (MC) generators to extract $m_t$. Hence the measured mass corresponds
in fact to the mass parameter in the MC. 
Work is proceeding to address the exact
difference between the measured mass and the pole mass, as presented, for
example, in Appendix~C of Ref.~\cite{top-mass-definitions}.  
One alternative way to address this problem is to extract $m_t$ from
a measurement of the \ttbar\ cross section~\cite{mass-from-xsec-measured}.
The {\dzero}
Collaboration has recently shown that the directly measured mass of the top quark
is closer to the pole mass extracted from a measurement of the \ttbar\
cross section than to an $\overline{\rm{MS}}$ mass
extracted in a similar way~\cite{mass-from-xsec-measured}. Hence, within
the precision of theory and data, the directly measured $m_t$
is best interpreted as the top-quark pole mass. 

CPT invariance predicts that a particle and its antiparticle
partner have the same mass. This has been checked for
the top quark by the \dzero, CDF, and CMS Collaborations, and the masses are found to hold within
the measurement uncertainties, with $\Delta m_t
= m_t - m_{\bar{t}} = 0.84 \pm
1.87$~GeV~\cite{top-mass-difference-dzero}, $\Delta
m_t = -3.3 \pm 1.7$~GeV~\cite{top-mass-difference-cdf}, and
$\Delta m_t = -0.44 \pm 0.53$~GeV~\cite{top-mass-difference-cms}, respectively. Thus, the top-quark 
mass combination in this paper assumes $m_t = m_{\bar{t}}$.


\vspace{-0.2in}
\subsection{Predictions based on the top-quark mass}
\label{implications-top-mass}

The internal consistency of the SM can be tested by
using different observables to predict the values
of others and then to compare the expectations with their measured
values. For example, the relation between the mass of the $W$~boson ($M_W$) and
$\sin^2 \theta_W$ (the electroweak mixing angle) includes higher-order
radiative corrections involving $m_t$, hence the smaller the uncertainty on the
measured $m_t$; the stronger is the test of consistency. 

\begin{figure}[!h!tb]
\vspace{-0.15in}
\includegraphics[width=3.7in,height= 3.5in]{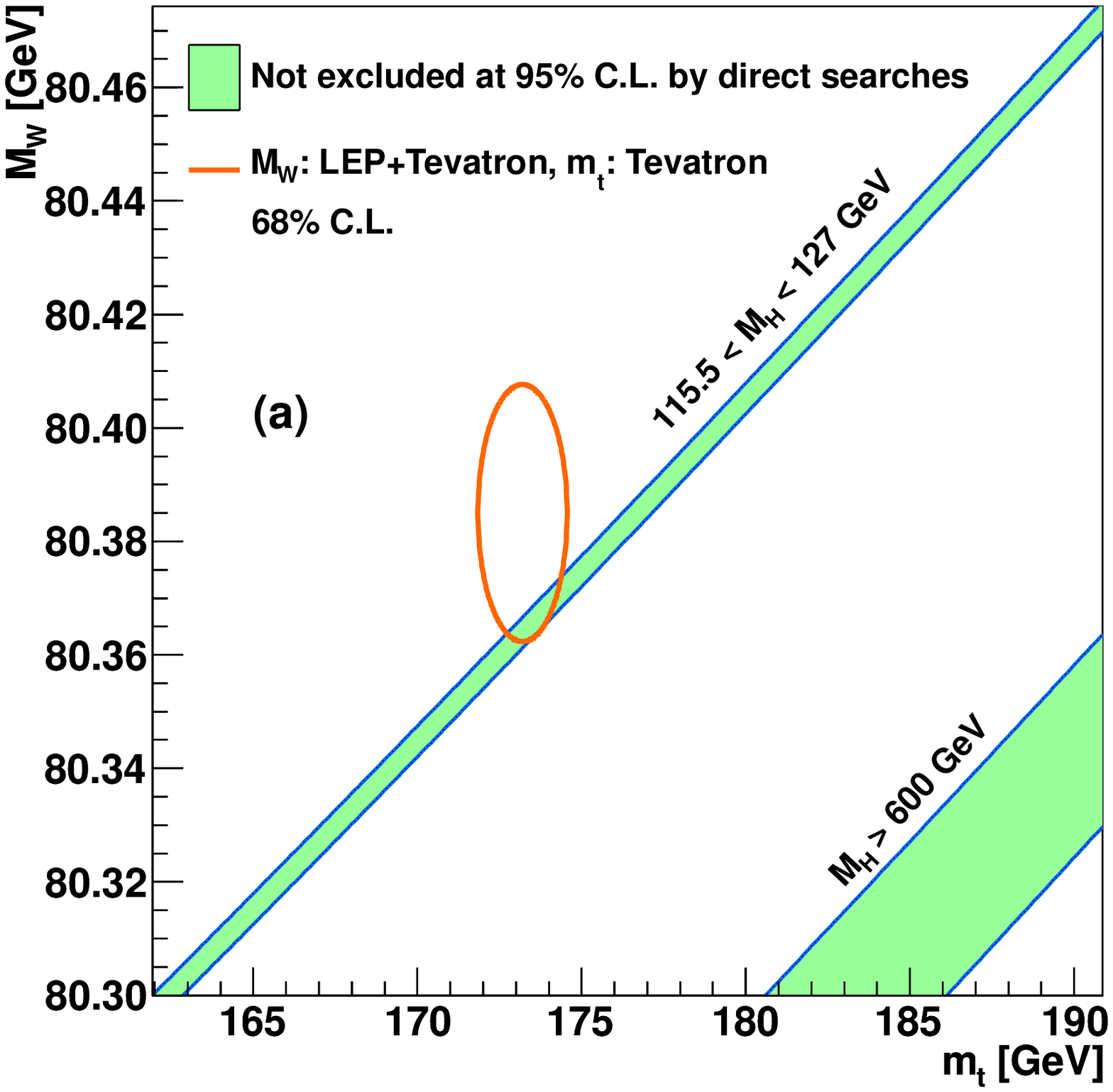}
\includegraphics[width=3.36in, height= 2.5in]{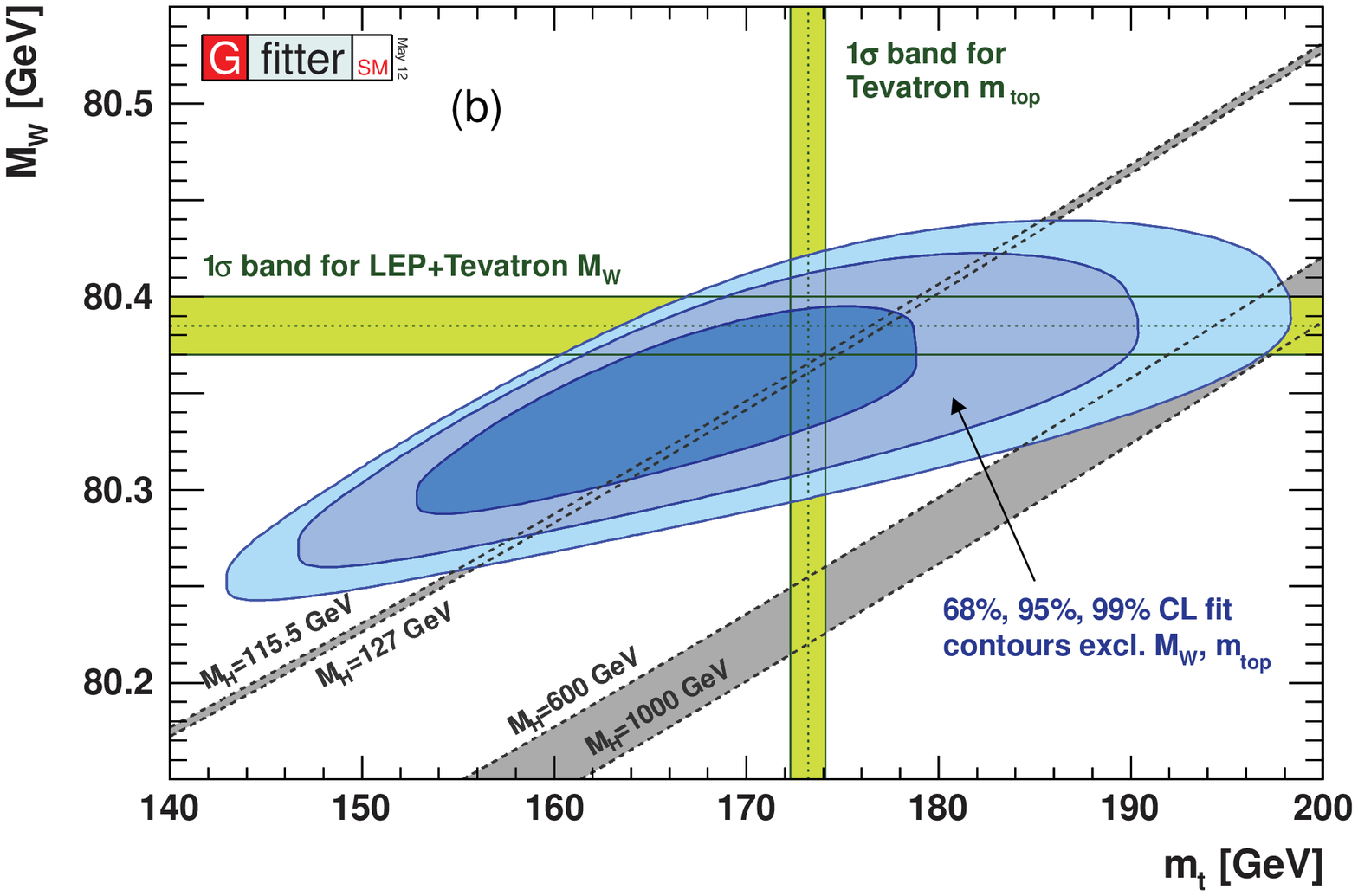}
\vspace{-0.1in}
\vspace{-0.1in}
\caption[mw_mt_mhiggs]{Constraints from LEP and
Tevatron measurements of $M_W$ and $m_t$ (Tevatron only) on
$M_H$ within the SM. 
The regions in the mass of the Higgs boson still allowed after the direct
searches at LEP, Tevatron and LHC are also shown.
From Ref.~\cite{gfitter}, the large countors (blue) indicate  
the constraints on the Higgs boson, from global fits to electroweak data without
including the direct measurements of $M_W$ and $m_t$ from the Tevatron.}
\label{mw-mt-mhiggs}
\vspace{-0.15in}
\end{figure}

Since 1997, the LEP Electroweak Working Group has used  the observed 
 top-quark and the $W$~boson masses and other precision electroweak variables 
to extract constrains on the
Higgs boson mass ($M_H$) in the SM~\cite{lepewwg}. This has been
extended to the minimal supersymmetric standard
model~\cite{heinemeyer}, and the 
GFITTER Collaboration has applied the technique to
set limits on a wide variety of theories beyond the SM~\cite{gfitter}. 
Figure~\ref{mw-mt-mhiggs}a shows the combined constraint attributable  to $M_W$ and $m_t$
(as of March 2012) on the Higgs boson mass.
Figure~\ref{mw-mt-mhiggs}b shows the constraint from $M_W$ and $m_t$ separately 
(as of March 2012) on the Higgs boson mass, and a global constraint originating 
from all the other electroweak variables, showing the importance of the $M_W$ and $m_t$ 
variables to constrain the Higgs boson mass.

\vspace{-0.2in}
\subsection{History of measurement of $m_t$}
\label{history}
\vspace{-0.1in}

Before 1995, global fits to electroweak data from the CERN and 
SLAC $e^+e^-$ colliders (LEP and SLC) and from other experiments
produced estimates of $m_t$ that
ranged from $\approx$~90~GeV to
$\approx$~190~GeV~\cite{mass-history-quigg}. At the time of the first observation 
of the top quark in 1995, the fits indicated a mass close to the current
Tevatron value of $m_t$, but with an uncertainty of
$\approx \pm 10\%$ and  an assumption of 300~GeV mass of the Higgs boson
 ~\cite{300GevHiggs}. 
CDF measured $m_t = 176 \pm8({\rm stat})
\pm10({\rm syst})$~GeV~\cite{top-observation-cdf} (total uncertainty of 7\%)
and {\dzero} measured $m_t = 199 ^{+19}_{-21}({\rm stat}) \pm22 ({\rm
syst})$~GeV~\cite{top-observation-dzero} (total uncertainty of 15\%).

\begin{figure*}[!h!tb]
\includegraphics[width=6.5in]{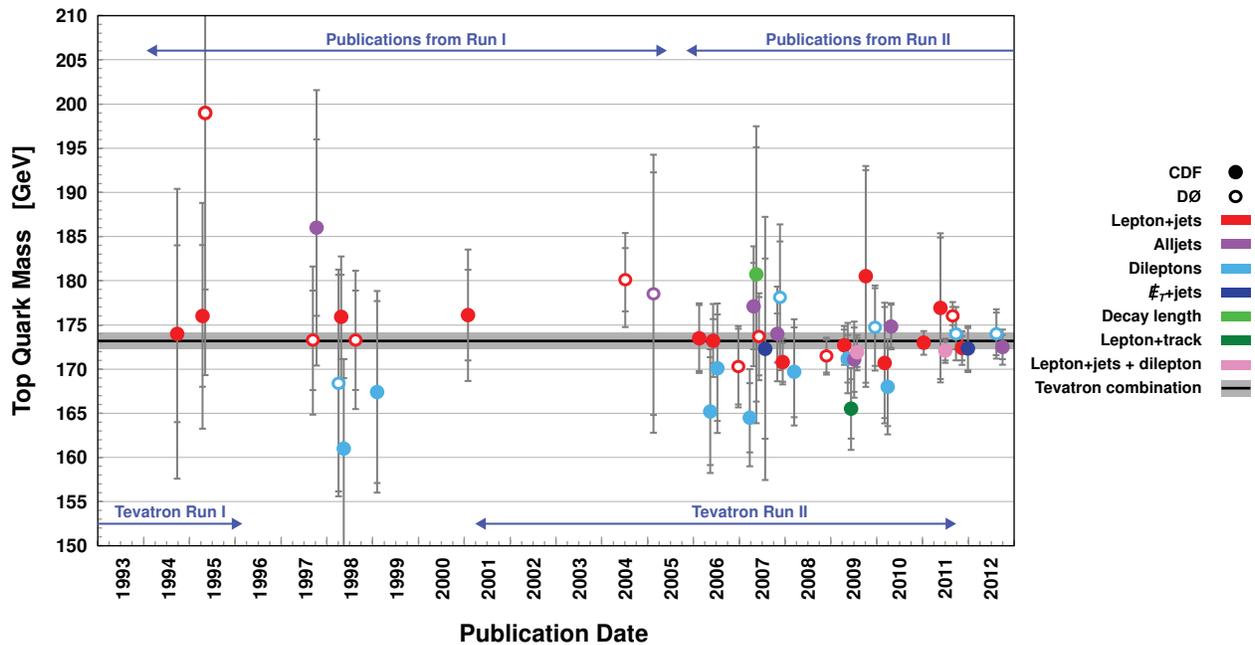}
\vspace{-0.1in}
\caption[top-mass-history]{The CDF and {\dzero} published direct measurements of the top-quark mass
as a function of time.}
\label{top_mass_history}
\end{figure*}

Since then, the CDF and {\dzero} Collaborations have
developed many novel measurement techniques and published nearly 50 journal papers
on their measurements of $m_t$. Recently, the CMS Collaboration at the
Large Hadron Collider (LHC) published a measurement using 102 dilepton
events~\cite{mass-measurement-cms} and finds $m_t = 175.5 \pm
4.6({\rm stat}) \pm4.6({\rm syst})$~GeV (total uncertainty of 3.7\%). The
ATLAS Collaboration at the LHC has submitted a measurement of 
$m_t = 174.5 \pm 0.6 \pm 2.3$~GeV (total uncertainty of 1.4\%) using nearly
12,000 lepton+jets events~\cite{mass-measurement-atlas}.  
The most precise measurements from the Tevatron in a single decay
channel use lepton+jets events, a matrix-element
method as introduced in Ref.~\cite{thesis-juan-estrada}, and an {\it in situ} calibration of the jet energy
scale. CDF's matrix-element measurement~\cite{mass-lepton+jets-runII-cdf}
uses 5.6~fb$^{-1}$ of integrated luminosity to find $m_t = 173.00 \pm 0.65({\rm stat})
\pm 1.06({\rm syst})$~GeV (total uncertainty of 0.72\%). {\dzero}'s
measurement~\cite{mass-lepton+jets-runII-dzero} uses 3.6~fb$^{-1}$ of
integrated luminosity to obtain $m_t = 174.94 \pm 0.83({\rm stat}) \pm 1.24({\rm
syst})$~GeV (total uncertainty of 0.85\%). Figure~\ref{top_mass_history}
shows the publication history of the direct measurements of $m_t$
at the Tevatron.


\vspace{-0.2in}
\subsection{Overview of mass measurements}
\label{mass-measurement-overview}
\vspace{-0.1in}

This paper reports on the combination of previously published
measurements of $m_t$.
Details of the analyses are therefore
not repeated as this information is available in recent 
reviews~\cite{top-mass-review}, as well as in the publications
of each of the results.
We will, however, summarize the basic techniques used for the measurements.

The cross section for $t\bar{t}$ production in
proton-antiproton ($p\bar{p}$) interactions at 1.96~TeV is
$\approx$~7.2~pb~\cite{top-pairs-xsec-expt,top-pairs-xsec-theory}. The mean
transverse momentum ($p_T$) of the $t\bar{t}$ system at parton level is $\approx$~20~GeV, which is attributed to
initial-state radiation (i.e., gluon emission).
The mean transverse momentum of the top quarks at parton level is 
$\approx$~95~GeV~\cite{top-quark-pT-dzero}. Top quarks
have a lifetime of $\approx$~$0.3\times10^{-24}$~s~\cite{top-lifetime-expt,top-lifetime-theory},
which is an order of magnitude smaller than the time scale for parton evolution 
and hadronization. 
Hence, when top quarks decay, they transfer their kinematic  characteristics 
to the $W$ boson and $b$ quark,
and the measured energy-momentum
four-vectors of the final-state particles can be used to reconstruct the mass of
the top quark, except for the presence of initial or final-state
radiation.

In alljets events, the four-vector of every jet emerging from quarks can be reconstructed,
but neutrinos emitted in semileptonic decays of $b$ quarks
and jet energy resolution effects will lead to lost energy.
In lepton+jets events, the momentum of the neutrino from the $W \to \ell \nu_\ell$
decay is not detected. The transverse component can be inferred
from the negative of the vector sum of all transverse momenta of particles 
detected in the calorimeter and muon detectors. We estimate the longitudinal momentum of $\nu_\ell$ 
by constraining the mass of the charged lepton and neutrino system to the 
world average value of $M_W$~\cite{particle-data-book}.
We also use $M_W$ to
choose the two light jets from $W \to q \bar{q'}$ decay, and we use that information 
for an {\it in situ} calibration of jet energies.
In dilepton events, the analysis is more complicated
because there are two final-state neutrinos from the leptonic decays of
both $W$~bosons. Therefore, the longitudinal and transverse-momentum 
components of the neutrinos cannot be determined without the application of
more sophisticated tools. These involve
assuming a value for $m_t$ to solve the event kinematics
and assigning a weight to each $m_t$ hypothesis
to determine the most likely value of $m_t$ consistent with the hypothesis that the event is a \ttbar\ event.

A major issue in $t\bar{t}$ final-state reconstruction is the  
correct mapping of the reconstructed objects to the partons from  the 
decays of the top quark and $W$~boson. The  problem arises because often  
the jet charge and flavor cannot be uniquely  determined. 
This creates combinatorial ambiguities in the $t\bar{t}$ event 
reconstruction that vary from 90 possible jet-to-parton assignments for the 
alljets final state to 2 in the dilepton channel. In the lepton+jets and dilepton 
final states, additional ambiguities may arise from 
multiple kinematical solutions for the longitudinal component of the 
neutrino momentum.

Two methods are used to measure the value of $m_t$. In the
first method, the reconstructed mass distribution in data, or a variable
correlated with $m_t$, such as the decay length of the $B$~hadron or
the transverse momentum of a lepton, is compared to template distributions composed 
of contributions from
background and simulation of $t\bar{t}$ events. One template is used to represent
background and another 
for each putative value of $m_t$. The second method
uses event probabilities based on the LO matrix element for the production of
$t\bar{t}$. For each event, a probability is
calculated as a function of $m_t$ that this event is
from $t\bar{t}$ production, as based on the corresponding production and decay matrix
element. Detector resolution is taken into account in the
calculation of these probabilities through transfer functions
that correlate parton-level energies and their measured values. The value of $m_t$
is then extracted from the joint probability calculated for all
selected events, based on the probability for signal and background (also defined through its matrix element). 
This method produces the most accurate results, but
the computations are time-consuming.


\vspace{-0.2in}
\subsection{Combination overview}
\label{combination-overview}
\vspace{-0.1in}

This paper describes the combination of statistically independent
top-quark mass measurements from the Fermilab Tevatron Collider.
Measurements are independent if they are based on different data sets,
e.g., from CDF and from {\dzero}, or from Tevatron Run~I (1992--1996)
and Run~II (2001--2011). They are also independent within one data set
if the event selections are designed to be
exclusive; i.e., no event can pass more than one category of
selections. At times, more than one measurement is published using
the same data and decay channel. 
In this situation,
the result with smallest overall uncertainty is chosen
for the combination. Twelve measurements are used in the combination described here,
eight from the CDF collaboration and four from {\dzero}. These
comprise five lepton+jets measurements (CDF and {\dzero},
Run~II and Run~I, and a CDF Run~II result based on the decay length of $B$ hadrons);
two alljets measurements (CDF Run~II and Run~I); four dilepton
measurements (CDF and {\dzero}, Run~II and Run~I); and a {\met}+jets
measurement (CDF Run~II). We combine these measurements using an
analytic method called the best linear unbiased estimator
(BLUE)~\cite{BLUE-method-1,BLUE-method-2,BLUE-method-3}. This technique
forms a linear combination of the separate unbiased mass measurements to
produce the best estimate of $m_t$ with the
smallest uncertainty. This procedure follows a series of 11
such mass combinations presented in~\cite{tevatron-combination-1,
tevatron-combination-2,tevatron-combination-3,tevatron-combination-4,
tevatron-combination-5,tevatron-combination-6,tevatron-combination-7,
tevatron-combination-8,tevatron-combination-9,tevatron-combination-10,
tevatron-combination-11}, updated each year since
2004 as new measurements of $m_t$ became available.
The combination presented here is the first to be published in a peer-reviewed journal.

%
%
%

\vspace{-0.1in}
\section{Inputs to the combination}
\label{combination-inputs}


\vspace{-0.1in}
\subsection{The independent mass measurements}
\label{twelve-measurements}
\vspace{-0.1in}

The mass measurements included in the combination are shown
in Table~\ref{input-masses}
~\cite{mass-lepton+jets-runII-cdf,mass-lepton+jets-runII-dzero,
       mass-lepton+jets-runI-cdf, mass-lepton+jets-runI-dzero,
       mass-alljets-runII-cdf,    mass-alljets-runI-cdf,
       mass-dilepton-runII-cdf,   mass-dilepton-runII-dzero,
       mass-dilepton-runI-cdf,    mass-dilepton-runI-dzero,
       mass-met+jets-runII-cdf,   mass-decaylength-runII-cdf}.
These 12 channels are chosen because they are statistically
independent, which maximizes the improvement in the combination, and
because enough information is available to separate
the components of systematic uncertainty for proper
treatment in the combination. 

The {\dzero} measurement from 2005 in the alljets channel (Run~I)~\cite{mass-alljets-runI-dzero} of
$m_t = 178.5 \pm 13.7({\rm stat}) \pm 7.7({\rm syst})$~GeV
(total uncertainty of 8.8\%) is not included in the combination because
some subcomponents of the systematic uncertainty are not available.

The CDF measurement from Run~II based on decay-length analysis~\cite{mass-decaylength-runII-cdf}
differs from the others in that it uses the
mean decay length of $B$~hadrons in $b$-tagged lepton+jets
events as the $m_t$-sensitive variable. It is independent of energy information in the calorimeter, and its main source of
systematic uncertainty is uncorrelated with the dominant ones from
the jet energy scale calibration in other measurements.
This measurement of $m_t$ is
essentially uncorrelated with the higher precision CDF result from the
lepton+jets channel.
The overlap between the data samples used for the decay-length method
and the lepton+jets sample has therefore no effect.

\begingroup
\squeezetable
\begin{table*}[!h!tb]
\caption[inputs]{Top-quark mass measurements used as input to determine the combined value of $m_t$ from the Tevatron
and the combined result.}
\label{input-masses}
\vspace{0.05in}
\begin{ruledtabular}
\begin{tabular}{lll.{1.1}.{5.0}.{2.0}.{3.2}c.{2.2}c.{1.2}.{1.2}c}
\multicolumn{13}{c}{   } \vspace{-0.07in} \\
Decay        & Tevatron & Experiment & \multicolumn{1}{c}{Integrated}  & \multicolumn{1}{c}{Number of}
                                     & \multicolumn{1}{c}{Background}  & \multicolumn{5}{c}{$m_t$}     & \multicolumn{1}{c}{Uncertainty} & Reference \\
channel      & period   &            & \multicolumn{1}{c}{luminosity}  & \multicolumn{1}{c}{events}      
                                     & \multicolumn{1}{c}{[\%]}        & \multicolumn{5}{c}{[GeV]}     & \multicolumn{1}{c}{on $m_t$}          &             \\
or method    &          &            & \multicolumn{1}{c}{[fb$^{-1}$]} & \multicolumn{1}{c}{ }    
                                     & \multicolumn{1}{c}{ }           & \multicolumn{5}{c}{ }         & \multicolumn{1}{c}{[\%]}        &             \\
[1.0ex] \hline \\ [-1.5ex]
Lepton+jets  & Run~II   & CDF        &  5.6  & 1087  &  17  &  173.00  & $\!\!\!\!\!\!\!\!\pm\!\!\!\!\!\!\!\!$ &  0.65 & $\!\!\!\!\!\!\!\!\pm\!\!\!\!\!\!\!\!$ &  1.06  &  0.72 & \cite{mass-lepton+jets-runII-cdf}  \\
Lepton+jets  & Run~II   & {\dzero}   &  3.6  &   615  &  27  &  174.94  & $\!\!\!\!\!\!\!\!\pm\!\!\!\!\!\!\!\!$ &  0.83 & $\!\!\!\!\!\!\!\!\pm\!\!\!\!\!\!\!\!$ &  1.24  &  0.85 & \cite{mass-lepton+jets-runII-dzero}\\
Lepton+jets  & Run~I    & CDF        &  0.1  &    76  &  54  &  176.1   & $\!\!\!\!\!\!\!\!\pm\!\!\!\!\!\!\!\!$ &  5.1  & $\!\!\!\!\!\!\!\!\pm\!\!\!\!\!\!\!\!$ &  5.3   &  4.2  & \cite{mass-lepton+jets-runI-cdf}   \\
Lepton+jets  & Run~I    & {\dzero}   &  0.1  &    22  &  22  &  180.1   & $\!\!\!\!\!\!\!\!\pm\!\!\!\!\!\!\!\!$ &  3.6  & $\!\!\!\!\!\!\!\!\pm\!\!\!\!\!\!\!\!$ &  3.9   &  2.9  & \cite{mass-lepton+jets-runI-dzero} \\
Alljets      & Run~II   & CDF        &  5.8  & 2856  &  71  &  172.47  & $\!\!\!\!\!\!\!\!\pm\!\!\!\!\!\!\!\!$ &  1.43 & $\!\!\!\!\!\!\!\!\pm\!\!\!\!\!\!\!\!$ &  1.40  &  1.2  & \cite{mass-alljets-runII-cdf}      \\
Alljets      & Run~I    & CDF        &  0.1  &   136  &  79  &  186.0   & $\!\!\!\!\!\!\!\!\pm\!\!\!\!\!\!\!\!$ & 10.0  & $\!\!\!\!\!\!\!\!\pm\!\!\!\!\!\!\!\!$ &  5.7   &  6.2  & \cite{mass-alljets-runI-cdf}       \\
Dileptons    & Run~II   & CDF        &  5.6  &   392  &  23  &  170.28  & $\!\!\!\!\!\!\!\!\pm\!\!\!\!\!\!\!\!$ &  1.95 & $\!\!\!\!\!\!\!\!\pm\!\!\!\!\!\!\!\!$ &  3.13  &  2.2  & \cite{mass-dilepton-runII-cdf}     \\
Dileptons    & Run~II   & {\dzero}   &  5.3  &   415  &  21  &  174.00  & $\!\!\!\!\!\!\!\!\pm\!\!\!\!\!\!\!\!$ &  2.36 & $\!\!\!\!\!\!\!\!\pm\!\!\!\!\!\!\!\!$ &  1.44  &  1.6  & \cite{mass-dilepton-runII-dzero}   \\
Dileptons    & Run~I    & CDF        &  0.1  &     8  &  16  &  167.4   & $\!\!\!\!\!\!\!\!\pm\!\!\!\!\!\!\!\!$ & 10.3  & $\!\!\!\!\!\!\!\!\pm\!\!\!\!\!\!\!\!$ &  4.9   &  6.8  & \cite{mass-dilepton-runI-cdf}      \\
Dileptons    & Run~I    & {\dzero}   &  0.1  &     6  &  25  &  168.4   & $\!\!\!\!\!\!\!\!\pm\!\!\!\!\!\!\!\!$ & 12.3  & $\!\!\!\!\!\!\!\!\pm\!\!\!\!\!\!\!\!$ &  3.6   &  7.6  & \cite{mass-dilepton-runI-dzero}    \\
{\met}+jets  & Run~II   & CDF        &  5.7  & 1432  &  32  &  172.32  & $\!\!\!\!\!\!\!\!\pm\!\!\!\!\!\!\!\!$ &  1.80 & $\!\!\!\!\!\!\!\!\pm\!\!\!\!\!\!\!\!$ &  1.82  &  1.5  & \cite{mass-met+jets-runII-cdf}     \\
Decay length & Run~II   & CDF        &  1.9  &   375  &  30  &  166.90  & $\!\!\!\!\!\!\!\!\pm\!\!\!\!\!\!\!\!$ &  9.00 & $\!\!\!\!\!\!\!\!\pm\!\!\!\!\!\!\!\!$ &  2.82  &  5.7  & \cite{mass-decaylength-runII-cdf}  \\ 
[1.0ex] \hline \\ [-1.5ex]
\multicolumn{3}{l}{Combination}
                                & \multicolumn{1}{c}{$\le5.8$} & 7420  &  44  &  173.18  & $\!\!\!\!\!\!\!\!\pm\!\!\!\!\!\!\!\!$ &  0.56 & $\!\!\!\!\!\!\!\!\pm\!\!\!\!\!\!\!\!$ &  0.75  &  0.54 & 
\vspace{0.02in}
\end{tabular}
\end{ruledtabular}
\end{table*}
\endgroup


\vspace{-0.15in}
\subsection{Data}
\label{data}
\vspace{-0.1in}

The data were collected with the CDF~\cite{detector-cdf} and
{\dzero}~\cite{detector-dzero,silicon-detector-dzero} detectors at the
Tevatron $p\bar{p}$ collider at Fermilab between 1992 and 2009. 
The Tevatron ``center-of-mass" energy was 1.8~TeV in Run~I from 1992 to 1996 and 1.96~TeV in Run~II from 2001. 
A silicon microstrip tracker around the beam pipe at the center of each
detector was used to reconstruct charged-particle tracks (only in Run~II at {\dzero}).
Tracks spatially matched to calorimeter jets are checked for originating from a secondary
vertex, or for evidence that they originate from decays of 
long-lived heavy-flavor hadrons containing $b$~quarks from the decay of top
quarks~\cite{detector-cdf,b-identification-dzero}. Electrons and
jets produce particle showers in the calorimeters, and the collected information is used
to measure their energies. 
Muons traverse the
calorimeters and outer muon detectors that are used to reconstruct
their tracks. Both CDF and {\dzero} have central axial magnetic
fields in the tracking region ({\dzero} only in Run~II), in which
the momenta of charged particles are determined from the curvature
of their tracks. The CDF magnet has a diameter of 3~m and extends 4.8~m
along the beam line, with a field strength of 1.4~T, and the {\dzero} magnet
has a diameter of 1.0~m and length of 2.7~m to fit inside
the Run~I calorimeter with a field strength of 2.0~T.
The CDF detector's larger tracking volume with a higher density of
measurements gives better transverse-momentum
resolution for charged-particle tracks. 
The transverse-momentum resolution is $\approx$ 3.5\% at CDF and $\approx$ 10~\% at {\dzero} for a muon with $p_T=50$~GeV.
The trigger and 
event-selection criteria depend on the \ttbar\ final states, details of which appear
in the publications listed in Table~\ref{input-masses}. The experiments
collected $\mathcal{O}(10^{14})$ hard collisions, from which 7420
events are selected because they have the characteristics expected for $t\bar{t}$ pairs, of which
$\approx$~56\% are expected to be true $t\bar{t}$ events.


\vspace{-0.15in}
\subsection{Models for \ttbar\ signal}
\label{signal-models}
\vspace{-0.1in}

The $t\bar{t}$ signal in Run~I was simulated using the LO
generator {\sc herwig}~\cite{herwig1} with
the MRSD$_0^{\prime}$~\cite{mrsd0p} and CTEQ4M~\cite{cteq4m}
parton distribution functions (PDF) used by CDF and {\dzero}, respectively. The {\sc herwig} generator
implements the hard-scattering processes $q\bar{q}{\rar}t\bar{t}$ and
$gg{\rar}t\bar{t}$, adding initial-state and
final-state radiation through leading-log QCD
evolution~\cite{dglap-evolution}. The top quark and $W$~boson in {\sc herwig} decay according
to the branching fractions listed by the Particle Data Group~\cite{particle-data-book}, and the
final-state partons are subsequently fragmented into jets. The MC events
are then processed through a fast simulation or a {\sc
geant} model~\cite{geant} of the detectors and then through event
reconstruction programs.

For the $t\bar{t}$ signal in Run~II, CDF uses {\sc pythia}~\cite{pythia2}
with the CTEQ5L~\cite{cteq5} PDF, and {\dzero}
uses the leading-log generator {\sc alpgen}~\cite{alpgen} with the
CTEQ6L1~\cite{cteq6} PDF and {\sc pythia} for parton showering.
{\sc alpgen} contains more tree-level graphs in higher-order $\alpha_s$ than {\sc pythia}.
{\sc alpgen} has parton-jet matching~\cite{parton-jet-matching}, which
avoids double counting of partons in overlapping regions of jet kinematics. CDF sets
the event generation factorization and renormalization scales $Q^2$
to $m_t^2 + p_{\perp}^2 + (P_1^2 + P_2^2)/2$, where $p_{\perp}$
is the transverse momentum characterizing the scattering process, and $P_1^2$ and
$P_2^2$ are the virtualities of the incoming partons.
{\dzero} sets the scales to $m_t^2 + \langle p_T^2 \rangle$, where $\langle p_T^2 \rangle$ is the
average of the square of transverse momentum of all other light partons produced in association with the
$t\bar{t}$ pair. The {\sc pythia} model treats each step of the
$t\bar{t}$ decay chain ($t{\rar}Wb$, $W{\rar}\ell\nu$ or
$q\bar{q}^{\prime}$) separately and does not preserve spin
correlations. {\sc alpgen} uses exact matrix elements for each step
and thereby correctly describes the spin information of the final-state partons. 
The fragments of the proton and antiproton or ``underlying event'' 
are added separately to each hard collision.
CDF uses the ``Tune~A'' settings~\cite{tuneA-cdf} in {\sc pythia} while {\dzero} uses
a modified version of the tune. Both collaborations use angular
ordering for modeling parton showering in {\sc pythia}, and not
$p_T$-ordered models. The underlying event is therefore
not interleaved with the parton showers as in models of color reconnection~\cite{color-reconnection-skands}.


\vspace{-0.15in}
\subsection{Background models}
\label{background-models}
\vspace{-0.1in}

In the lepton+jets channel, the dominant background is from $W$+jets production.
Smaller contributions arise from multijet events, $Z$+jets, single top-quark 
($tqb$ and $tb$), and diboson production ($WW$, $WZ$, and
$ZZ$). The alljets channel has mainly multijet events as background. The
largest background in the dilepton channel is from $Z$+jets events, which include
Drell-Yan production. Backgrounds from diboson production and from events with jets identified as leptons 
are very small in the dilepton channel. 
The {\met}+jets channel has multijet events and $W$+jets as main backgrounds. 

In all
channels contributions from multijet events are modeled using data.
Most other background sources are modeled through MC simulation.
In Run~I, both collaborations used {\sc vecbos}~\cite{vecbos} to model
$W$+jets events. {\sc vecbos} is a precursor of {\sc alpgen} and
provides one of the first models of events with many high-momentum
final-state partons. {\sc pythia} was used to model $Z$+jets, Drell-Yan, and
diboson processes. Background from events with a single top quark was negligible. In Run~II,
both collaborations used {\sc
alpgen} for the simulation of the $W$+jets background. The treatment of
heavy-flavor jets is implemented more accurately in {\sc alpgen},
and parton-jet matching also improves the
simulation. For the $Z$+jets background, CDF uses {\sc pythia} and
{\dzero} uses {\sc alpgen}. For dibosons, both collaborations use {\sc
pythia}. Processes with a single top quark are modeled by CDF using {\sc
madevent}~\cite{madevent} (based on {\sc madgraph}~\cite{madgraph}), and by 
{\dzero} with {\sc singletop}~\cite{singletop}
(based on {\sc comphep}~\cite{comphep}).

The uncertainty in the description of the $W$+jets background has three main components: (i) the uncertainty on the scale $Q^2$,
which affects both the overall normalization and the differential jet distributions in pseudorapidity 
$\eta$~\cite{pseudorapidity}
and $p_T$; (ii) the uncertainty in the correction for flavor content of jets to higher
order; and (iii) the limitation in the MC model we are using to reproduce the jet $p_T$ and $\eta$ distributions in data
at low $p_T$ and large $|\eta|$.


\vspace{-0.15in}
\subsection{Jet properties}
\label{all-about-jets}
\vspace{-0.14in}

After the top quarks decay, the final-state quarks and
gluons hadronize to produce multiple charged and neutral particles
that traverse the central tracking systems into the calorimeters,
where they produce many lower-momentum particles through interactions in
the absorbers of the calorimeters.
The observed particles tend to cluster in jets that can be assigned to the initial partons.
For jet reconstruction, the CDF
Collaboration uses a clustering algorithm in
$(\eta,\phi)$ space~\cite{jet-algorithm-cdf} with a cone radius of
\begin{align*}
{\rm CDF} \hspace{0.4in} {\cal R} =
\sqrt{(\Delta\eta)^2 + (\Delta\phi)^2} = 0.4,
\end{align*}
\noindent where $\phi$ is the azimuthal angle around the beamline,
$\eta$ is the pseudorapidity, and $\Delta \eta$ or $\Delta \phi$
are the widths of the cone.
{\dzero} uses a midpoint iterative seed-based cone algorithm in
$(y,\phi)$ space~\cite{jet-algorithm-dzero} with a radius defined by
\begin{align*}
{\rm \dzero} \hspace{0.4in} {\cal R} =
\sqrt{(\Delta y)^2+(\Delta\phi)^2} = 0.5,
\end{align*}
\noindent where the rapidity $y = 1/2 \ln \left( \left(E+p_L\right) /
\left(E-p_L\right) \right)$, $E$ is the jet energy, and $p_L$ is its
longitudinal momentum component. 

The jet energy resolution in the central region ($|\eta|<1$) is approximately
the same for CDF and {\dzero}; for CDF it is $\sigma(E_T)/E_T = 50\%/\sqrt{E_T (GeV)} \oplus 3\%$.
For jets in the forward region, however, the energy resolution at {\dzero} is similar
to that in the central region, while at CDF it is not as good [$\sigma(E_T)/E_T = 70\%/\sqrt{E_T (GeV)} \oplus 4\%$].
CDF's calorimeter covers $|\eta| < 3.8$ whereas
{\dzero}'s calorimeter covers $|\eta| < 4.2$. The {\dzero} calorimeter is more homogeneous,
so that the imbalance in transverse momentum (see Sec.~\ref{everyting-else}) 
usually has better resolution at {\dzero}. 
For both CDF and {\dzero}, to reject jets with mismeasured energy, selections on energy deposition
 are required when clustering the energy from the calorimeter cells into jets.
When a muon is reconstructed within the jet cone, a correction is applied
to the jet energy to account for the muon and its associated
neutrino assumed to arise from heavy-quark decay.

Jet energy scale calibrations are applied after jet reconstruction.
CDF calibrates the
transverse momentum using test-beam data and single-particle simulated
events and corrects the jet energy to the parton level. Consequently, CDF does
not calibrate the jet energy scale in MC events.
{\dzero} calibrates the energy using photon+jets and two-jet data and
calibrates jets in data as well as in MC to the observed particle
level.
Particle jets are clustered from stable particles after fragmentation, 
including particles from the underlying event, but excluding undetected 
energy from muons and neutrinos.

CDF's jet calibration~\cite{jet-energy-scale-cdf} applies two scale
factors and three offsets to convert the measured transverse momentum
of a jet to that of the parton that initiated the jet. {\dzero}'s jet
calibration~\cite{jet-energy-scale-dzero} applies three scale factors
and one offset to the jet energy to convert to the particle jet energy scale. The
calibrations are expressed as follows:
\begin{align*}
{\rm CDF} \hspace{0.25in} p_T^{\rm parton} & =
\frac{p_T^{\rm jet} \; R_{\rm rel} - C_{\rm MI}}{R_{\rm abs}}
- C_{\rm UE} + C_{\rm OC}, \\
{\rm \dzero} \hspace{0.2in} E^{\rm particle} & =
\frac{E^{\rm jet} - C_{\rm MI,UE}}{R_{\rm abs} \; R_{\rm rel} \;
F_{\rm OC}}.
\end{align*}
The absolute response $R_{\rm abs}$ corrects for energy lost
in uninstrumented regions between calorimeter modules, for differences
between electromagnetically and hadronically interacting particles,
as well as for module-to-module
irregularities. The relative response $R_{\rm rel}$ is a scale factor
that corrects forward relative to central jets and $C_{\rm MI}$ is a
correction for multiple interactions in the same bunch crossing.
The function $C_{\rm UE}$ is a correction for the jet energy added from the underlying
event. {\dzero} has one offset correction, $C_{\rm MI,UE}$, which
includes the effects of multiple interactions, the underlying event,
noise from radioactive decays of the uranium absorber, and the effect
of collisions from previous bunch crossings (pileup). The functions $C_{\rm OC}$
and $F_{\rm OC}$ are corrections for shower particles scattered in or out of the cone of radius ${\cal R}$. 
CDF's correction accounts for MC modeling that affects how the
parton energy is translated into particle jet energy, whereas
{\dzero}'s correction accounts for a detector effect caused by the finite
cell size in the calorimeter coupled with the cone size for the
jet algorithm. The combined jet energy scale corrections increase the
measured jet energies by about 20\%--50\%, depending on $p_T$ and
$\eta$.

The overall uncertainties on the jet energy scale corrections vary
from about 2.7\% for CDF and 1.1\% for {\dzero} for central
jets of transverse energy of 100~GeV to 3.3\% for CDF and 2.2\% for {\dzero} 
for forward jets. Central jets 
of 25~GeV have correction uncertainties of 5.9\% for CDF and 1.4\%
for {\dzero}. For both experiments, the uncertainty on the corrections for absolute response
$R_{\rm abs}$ dominate these uncertainties.

At {\dzero}, the jet energy resolution in data is inferior than predicted
by the detector simulation.  Therefore, the energies of MC jets
are smeared so that the resulting resolution in MC matches that in data.  
Similarly, the reconstruction efficiency for jets
in data is lower than is predicted by the detector simulation,
so an appropriate fraction of MC jets are randomly removed.
Both effects are corrected
for as functions of jet $p_T$ and pseudorapidity.

{\dzero} Run~II analyses include an energy correction to simulated jets
that depends on jet flavor. There are corrections for $b$~jets,
other-quark flavor jets ($u$, $d$, $s$, and $c$), and gluon jets implemented in both
the lepton+jets and dilepton analyses. Such corrections
refine the simulation by improving the matching of jet energies in MC to
data. The differences arise from the varying electromagnetic fractions
and widths of the jets. The corrections depend on jet
transverse energy and pseudorapidity and range from $-6\%$ to
$+2\%$~\cite{mass-lepton+jets-runII-dzero}.

Both collaborations perform an {\em in situ} jet energy scale
calibration
in lepton+jets events
for the matrix-element
mass extraction of $m_t$, and in CDF's alljets and {\met}+jets
measurements of $m_t$. The invariant mass of the two jets is constrained to
a Breit-Wigner distribution for the $W \to q \bar{q'}$ decay, set to
the world average value for the $W$-boson mass~\cite{particle-data-book}. 
The energies of all jets in the
event are then rescaled to complete this calibration.


\vspace{-0.15in}
\subsection{$b$-quark jet properties}
\label{a-bit-about-bs}
\vspace{-0.14in}

To separate top-quark events from background and to decrease the
ambiguity in jet-to-parton matching, it is important to identify
$b$-quark jets. Every $t\bar{t}$ event has two
$b$~jets, whereas such jets are rare in
background. As $B$~hadrons have
a mean lifetime of $\approx 10^{-12}$~s, $b$~jets can be tagged through
secondary vertices of the $B$ decay a few mm away from the primary 
$p\bar{p}$ interaction. 
CDF's $b$-tagging algorithm uses the significance of the displacement of the
secondary vertex in the transverse $(r,\phi)$ plane for the lepton+jets and
{\met}+jets channels~\cite{detector-cdf}, as well as a jet-probability
algorithm for {\met}+jets events~\cite{cdf-jetprob}. One parameter defines the
significance of the separation of the primary and secondary vertices for 
events with one and two $b$~jets.
For jets that are within the acceptance of the silicon microstrip
tracker (i.e., ``taggable'' jets), this algorithm identifies
50\% of real $b$ jets and 9\% of real charm jets, while falsely
tagging 1\% of light jets.
{\dzero} tags jets by combining nine track and
secondary-vertex-related variables using a neural 
network~\cite{b-identification-dzero}. 
For jets within the acceptance of the silicon microstrip detector,
this yields efficiencies of 65\% and 20\% for real $b$ and charm jets,
respectively, while falsely tagging 3\% of light jets.

To identify heavy-flavor jets in data and in MC events, the tagging
algorithm is applied by CDF and {\dzero} directly to the jets, except for
simulated $W$+light jets events, where CDF uses tag-rate functions measured
in multijet data, since the rate for directly-tagged MC events is
very low. After applying direct tagging to $b$ and $c$ jets in MC events, {\dzero}
corrects the tagging efficiencies to match those observed in data by
randomly dropping the tagging of 13\% of such jets.
For light-flavor jets, {\dzero} assigns a per jet mistag weight.


\vspace{-0.15in}
\subsection{Properties of other event observables}
\label{everyting-else}
\vspace{-0.14in}

The uncertainty on $m_t$ depends not only on an accurate measurement of
jet energies and proper assignment of flavor, but also on the
reconstruction and calibration of the other elements of the event,
including electrons, muons, and the imbalance in transverse momentum,
taking into account the presence of any simultaneous $p\bar{p}$ interactions in the same bunch crossing.

The mean number of $p\bar{p}$ collisions per bunch crossing is $\approx 2$ 
in Run~I and $\approx 5$ in Run~II. Such additional collisions affect
the observed characteristics of the hard scatter of interest and must be included in the
MC simulation. These extra collisions result mostly in the production of
low-$p_T$ particles. CDF simulates such additional interactions using
the {\sc pythia} model of minimum-bias events and overlays them onto the
hard scatters using a Poisson mean appropriate
to the instantaneous luminosity of the data. In a similar manner {\dzero} overlays
randomly triggered data events with the same luminosity profile as the data onto the MC simulated events.

Electrons are identified by matching clusters of energy deposited
in the electromagnetic layers of the calorimeters with tracks 
that point from the primary collision vertex to the clusters.
The spatial shapes of the
showers must agree with those expected for electrons, as studied in
test-beam data. 
The energy of an electron is determined as a  combination of the
total energy of the cluster and the momentum measured from
the curvature of the matching track.
The reconstruction efficiency is determined using
$Z \rightarrow ee$ data by identifying one tight charged lepton a as tag and using the
other charged lepton as a probe (tag-and-probe method).
The electron energy is also recalibrated using such $Z$ events.

Muons are reconstructed from a central track  and matched 
to a track in the outer muon
chambers. In {\dzero}, both the inner and outer trajectories pass through
magnetic fields, and so the transverse momenta of the two are therefore required to
match. The reconstruction efficiency and
calibration of $p_T$ are determined using a tag-and-probe method applied on $J/\psi \to \mu\mu$ and 
$Z \to \mu\mu$ events in a manner similar to that used for electrons.

As indicated above, all \ttbar\ decay channels except for alljets
events have a large \met.
All jet energy calibration corrections are also propagated
to \met\ in each event.

%
%
%

\section{Combination of mass measurements}
\label{combination}


\vspace{-0.1in}
\subsection{BLUE combination method}
\label{BLUE-method}
\vspace{-0.1in}

The basic idea of the technique, called the  best linear unbiased estimator (BLUE) method~\cite{BLUE-method-1,BLUE-method-2,BLUE-method-3},
used to obtain the combined mass $m_t^{\rm comb}$, an ``estimator'' of the true mass
$m_t^{\rm true}$, 
is to calculate a linear weighted sum of the results from separate measurements:
\begin{align}
m_t^{\rm comb} & = \sum_{i=1}^{12} w_i \; m_t^i.
\end{align}
The $m_t^i$ are the 12 CDF and {\dzero} measurements $i$ of $m_t$ and
\begin{align}
\sum_{i=1}^{12} w_i = 1.
\end{align}
The weights are determined using the value of $m_t^{\rm comb}$
that minimizes the squared difference relative to the unknown true value $m_t^{\rm true}$:
\begin{multline}
\left(m_t^{\rm comb} - m_t^{\rm true}\right)^2 = \\
      {\rm Variance} \! \left(m_t^{\rm comb}\right)
      + \left[{\rm Bias} \! \left(m_t^{\rm comb}\right)\right]^2,
\end{multline}
where the two terms represent the weighted variance and bias in the 12 input $m_t$ values with
\begin{align}
{\rm Variance} \! \left(m_t^{\rm comb}\right) = 
   & \sum_{i=1}^{12} w_i^2 \; {\rm Variance} \! \left(m_t^i\right),
\end{align}
and
\begin{align}
{\rm Variance} \! \left(m_t^i\right) =
   & \left[\sigma \! \left(m_t^i\right)\right]^2,
\end{align}
where $\sigma \! \left(m_t^i\right)$ are the uncertainties on the 12 input
values given in Table~\ref{input-masses}.

\vspace{0.2in}
On average, we expect the input mass measurements to be unbiased, and
we therefore assume
\begin{align}
{\rm Bias} \! \left(m_t^{\rm comb}\right) =
\sum_{i=1}^{12} w_i \; {\rm Bias} \! \left(m_t^i\right) = 0.
\end{align}
Equation~(3) shows that the BLUE method defines the best estimate
through a minimization of the variance of $m_t$ for an assumed
unbiased set of measurements.
The minimum corresponds to setting the weights to
\begin{align}
w_i   & = \frac{1/{\rm Variance} \! \left({m_t^i}\right)}
               {\sum_{i=1}^{12} 1/{\rm Variance} \! \left(m_t^i\right)}
\end{align}
for uncorrelated input values.
Since the input $m_t$ values are correlated,
the variance in Eq.~(4) has to be replaced with a
covariance matrix:
\begin{multline}
{\rm Variance} \! \left(m_t^{\rm comb}\right) = \\
      \sum_{i=1}^{12} \sum_{j=1}^{12} w_i \; w_j \;
       {\rm Covariance} \! \left(m_t^i, m_t^j\right),
\end{multline}
\noindent which is defined as
\begin{multline}
{\rm Covariance} \! \left(m_t^i,m_t^j\right) = \\
     \left[\sigma \! \left(m_t^i \; m_t^j\right)\right]^2
     - \sigma \! \Bigl(m_t^i\Bigr) \sigma \! \left(m_t^j\right).
\end{multline}
Minimizing Eq.~(3) yields
\begin{align}
w_i   = \frac{\sum_{j=1}^{12}
             {\rm Covariance}^{-1} \! \left(m_t^i,m_t^j\right)}
             {\sum_{i=1}^{12} \sum_{j=1}^{12}
             {\rm Covariance}^{-1} \! \left(m_t^i,m_t^j\right)},
\end{align}
\noindent where ${\rm Covariance}^{-1} \! \left(m_t^i,m_t^j\right)$
are the elements of the inverse of the covariance matrix (also
known as the error matrix), and
\begin{multline}
   {\rm Covariance} \! \left(m_t^i,m_t^j\right) = \\
     {\rm Correlation} \! \left(m_t^i,m_t^j\right)
        \sigma \! \Bigl(m_t^i\Bigr) \sigma \! \left(m_t^j\right)
\end{multline}
with ${\rm Correlation} \! \left(m_t^i,m_t^j\right)$ the correlation
coefficient between $m_t^i$ and $m_t^j$.
The following sections show how the correlation matrix is derived by
examining the uncertainty components and their individual
correlations.


\vspace{-0.2in}
\subsection{Measurement uncertainties}
\label{measurement-uncertainties}
\vspace{-0.15in}

The uncertainty on any $m_t$ measurement has a statistical component from
the limited number of events available for the measurement and a
systematic component from the uncertainties assigned to the
calibration of input quantities, to the model of the signal, and to the
calibration of the mass extraction method. Since the first measurements of
$m_t$~\cite{top-observation-cdf,top-observation-dzero}, the systematic component
has been slightly larger than the statistical one. As more data
became available, the statistical uncertainties on $m_t$ improved as did
the calibrations of systematic uncertainty, and the two components therefore improved together.

The systematic uncertainty on each $m_t$ measurement in this combination is divided
into 14 parts. Some of them have origin in only one
source whereas, others include several 
related sources of uncertainties. 
For the latter the patterns of correlation among different channels,
Tevatron Run~I and Run~II, or experiments are the same for all sources
included in these systematic components. The uncertainty on jet energy scale (JES),
on the other hand, is split into seven components, which do not appply to 
all measurements, given the
significantly different approaches to jet energy calibration between CDF and
{\dzero} and the change in the {\dzero} procedure between Run~I and
Run~II.

\begingroup
\squeezetable
\begin{table*}[!h!tb]
\caption[uncertainties]{The uncertainty in GeV from each component for the
12  measurements of $m_t$ and the resulting Tevatron
combination. The total uncertainties are obtained by
adding the components in quadrature.  The entries ``n/a'' stand for ``not
applicable'' and ``n/e'' for ``not evaluated.'' The nonevaluated uncertainties were not considered
as significant sources of uncertainty for Run~I measurements.}
\label{uncertainties}
\vspace{0.05in}
\begin{ruledtabular}
\begin{tabular}{lll|.{1.2}.{1.2}.{1.2}.{1.1}.{1.2}.{1.2}.{1.2}|.{1.2}.{1.2}.{1.2}.{1.2}.{1.2}.{1.2}.{1.2}|.{2.2}.{1.2}.{1.2}.{2.2}}
 &  & 
& \multicolumn{1}{c}{\begin{sideways}Light-jet response (1)\end{sideways}}
& \multicolumn{1}{c}{\begin{sideways}Light-jet response (2)\end{sideways}}
& \multicolumn{1}{c}{\begin{sideways}Out-of-cone correction\end{sideways}}
& \multicolumn{1}{c}{\begin{sideways}Offset\end{sideways}}
& \multicolumn{1}{c}{\begin{sideways}Model for $b$ jets\end{sideways}}
& \multicolumn{1}{c}{\begin{sideways}Response to $b/q/g$ jets\end{sideways}}
& \multicolumn{1}{c|}{\begin{sideways}{\it In situ} light-jet calibration\end{sideways}}
& \multicolumn{1}{c}{\begin{sideways}Jet modeling\end{sideways}}
& \multicolumn{1}{c}{\begin{sideways}Lepton modeling\end{sideways}}
& \multicolumn{1}{c}{\begin{sideways}Signal modeling\end{sideways}}
& \multicolumn{1}{c}{\begin{sideways}Multiple interactions model\end{sideways}}
& \multicolumn{1}{c}{\begin{sideways}Background from theory\end{sideways}}
& \multicolumn{1}{c}{\begin{sideways}Background based on data~~\end{sideways}}
& \multicolumn{1}{c|}{\begin{sideways}Calibration method\end{sideways}}
& \multicolumn{1}{c}{\begin{sideways}{\bf Statistical uncertainty}\end{sideways}}
& \multicolumn{1}{c}{\begin{sideways}{\bf Total JES uncertainty}\end{sideways}}
& \multicolumn{1}{c}{\begin{sideways}{\bf Other systematic uncertainty}\end{sideways}}
& \multicolumn{1}{c}{\begin{sideways}{\bf Total uncertainty}\end{sideways}} \\
[1.0ex] \hline \\ [-1.5ex]
Channel & Run & Exp. & \multicolumn{7}{c|}{Jet energy scale systematics} & \multicolumn{7}{c|}{Other systematics} & & & & \\
[1.0ex] \hline \\ [-1.5ex]
Lepton+jets  & II   & CDF~~        & 0.41 & 0.01 & 0.27 & \nap & 0.23 & 0.13 & 0.58 & 0.00 & 0.14 & 0.56 & 0.10 & 0.27 & 0.06 & 0.10 &  0.65 & 0.80 & 0.67 &  1.23 \\
Lepton+jets  & II   & {\dzero}     & \nap & 0.63 & \nap & \nap & 0.07 & 0.26 & 0.46 & 0.36 & 0.18 & 0.77 & 0.05 & 0.19 & 0.23 & 0.16 &  0.83 & 0.83 & 0.94 &  1.50 \\
Lepton+jets  & I    & CDF          & 3.4  & 0.7  & 2.7  & \nap & 0.6  & \nev & \multicolumn{1}{c|}{n/a} & \nev & \nev & 2.7  & \nev & 1.3  & \nev & 0.0  &  5.1  & 4.4  & 2.8  &  7.3  \\
Lepton+jets  & I    & {\dzero}     & \nap & 2.5  & 2.0  & 1.3  & 0.7  & \nev & \multicolumn{1}{c|}{n/a}  & \nev & \nev & 1.3  & \nev & 1.0  & \nev & 0.6  &  3.6  & 3.5  & 1.6  &  5.3  \\
Alljets      & II   & CDF          & 0.38 & 0.04 & 0.24 & \nap & 0.15 & 0.03 & 0.95 & 0.00 & \nap & 0.64 & 0.08 & 0.00 & 0.56 & 0.38 &  1.43 & 1.06 & 0.91 &  2.00 \\
Alljets      & I    & CDF          & 4.0  & 0.3  & 3.0  & \nap & 0.6  & \nev & \multicolumn{1}{c|}{n/a}  & \nev & \nap & 2.1  & \nev & 1.7  & \nev & 0.6  & 10.0  & 5.0  & 2.6  & 11.5  \\
Dileptons    & II   & CDF          & 2.01 & 0.58 & 2.13 & \nap & 0.33 & 0.14 & \multicolumn{1}{c|}{n/a}  & 0.00 & 0.27 & 0.80 & 0.23 & 0.24 & 0.14 & 0.12 &  1.95 & 3.01 & 0.88 &  3.69 \\
Dileptons    & II   & {\dzero}     & \nap & 0.56 & \nap & \nap & 0.20 & 0.40 & 0.55 & 0.50 & 0.35 & 0.86 & 0.00 & 0.00 & 0.20 & 0.51 &  2.36 & 0.90 & 1.11 &  2.76 \\
Dileptons    & I    & CDF          & 2.7  & 0.6  & 2.6  & \nap & 0.8  & \nev & \multicolumn{1}{c|}{n/a}  & \nev & \nev & 3.0  & \nev & 0.3  & \nev & 0.7  & 10.3  & 3.9  & 3.0  & 11.4  \\
Dileptons    & I    & {\dzero}     & \nap & 1.1  & 2.0  & 1.3  & 0.7  & \nev & \multicolumn{1}{c|}{n/a}  & \nev & \nev & 1.9  & \nev & 1.1  & \nev & 1.1  & 12.3  & 2.7  & 2.3  & 12.8  \\
{\met}+jets  & II   & CDF          & 0.45 & 0.05 & 0.20 & \nap & 0.00 & 0.12 & 1.54 & 0.00 & \nap & 0.78 & 0.16 & 0.00 & 0.12 & 0.14 &  1.80 & 1.64 & 0.78 &  2.56 \\
Decay length & II   & CDF          & 0.24 & 0.06 & \nap & \nap & 0.15 & \nev & \multicolumn{1}{c|}{n/a}  & 0.00 & \nap & 0.90 & 0.00 & 0.80 & 0.20 & 2.50 &  9.00 & 0.25 & 2.80 &  9.43 \\ 
[1.0ex] \hline \\ [-1.5ex]
\multicolumn{3}{l|}{Tevatron combination}
                                       & 0.12 & 0.19 & 0.04 & 0.00 & 0.15 & 0.12 & 0.39 & 0.11 & 0.10 & 0.51 & 0.00 & 0.14 & 0.11 & 0.09 & 0.56 & 0.49 & 0.57 & 0.94
\vspace{0.02in}
\end{tabular}
\end{ruledtabular}
\end{table*}
\endgroup

Table~\ref{uncertainties} gives the uncertainty of each of the 12 top-quark 
mass measurements for the different contributions to uncertainty
and their effect on the final combination. The components of
uncertainty are defined in the following and can be classified as
uncertainties in detector response (jet energy scale, jet and lepton modeling),
uncertainties from modeling signal and background (signal modeling, multiple interactions model,
background estimated from theory, and background based on data), uncertainties from
method of mass extraction, and statistical uncertainties.
A detailed description of the methods to evaluate these systematic uncertainties is presented in the Appendix.

\vspace{0.1in}
\noindent {\bf 1.~~Jet energy scale}\\*[0.05in]
\noindent {\bf 1.1~~Light-jet response (1)}\\*
\label{rJES}
\indent One subcomponent of the uncertainty in JES covers the
absolute calibration for CDF's Run~I and Run~II
measurements. It also includes small contributions from the
uncertainties associated with modeling multiple interactions within a
single bunch crossing and corrections for the underlying event.

\vspace{0.08in}
\noindent {\bf 1.2~~Light-jet response (2)}\\*
\label{dJES}
\indent Another subcomponent of this uncertainty includes
{\dzero}'s Run~I and Run~II calibrations of absolute response
(energy dependent), the relative response ($\eta$ dependent),
and the out-of-cone showering correction that is a detector effect. This
uncertainty term for CDF includes only the small relative response
calibration ($\eta$ dependent) for Run~I and Run~II.

\vspace{0.08in}
\noindent {\bf 1.3~~Out-of-cone correction}\\*
\label{cJES}
\indent This subcomponent of the JES uncertainty quantifies
the out-of-cone showering corrections to the MC showers
for all of CDF's and for {\dzero}'s Run~I measurements that are obtained by varying
the model for light-quark fragmentation.

\vspace{0.08in}
\noindent {\bf 1.4~~Offset}\\*
\label{UN/MI}
\indent This subcomponent originates from the
offset in {\dzero}'s Run~I calibration, which corrects for noise from
uranium decay, pileup from previous collisions, and for
multiple interactions and the model for the underlying event. In Run~I, the
uncertainties are large, but in Run~II, owing to the smaller
integration time for calorimeter electronics, they are negligible. CDF's
calorimeter does not have the same sources of noise and sensitivity to
pileup as {\dzero}, so CDF measurements do not have this term.

\vspace{0.08in}
\noindent {\bf 1.5~~Model for $\mathbi{b}$ jets}\\*
\label{bJES}
\indent This subcomponent comes from
the uncertainty on the semileptonic branching fraction in $b$
decays and from differences between two models of $b$-jet
hadronization.

\vspace{0.08in}
\noindent {\bf 1.6~~Response to $\mathbi{b/q/g}$ jets}\\*
\label{aJES}
\indent This subcomponent accounts
for the difference in the electromagnetic versus hadronic response of
$b$~jets, light-quark jets, and gluon jets.
CDF corrects for jet flavor as part of the main
calibration, and defines the uncertainty based on
the remaining difference in response between $b$~jets and light-flavor
jets, whereas {\dzero} corrects the response for $b$, light-quark
($u$, $d$, $s$, and $c$), and gluon jets as a function of
jet $p_T$ and $\eta$.

\vspace{0.1in}
\noindent {\bf 1.7~~{{\em In situ} light-jet calibration}}\\*
\label{iJES}
\indent The last part of the uncertainty in the jet energy scale
is from the {\it in situ} calibration of $m_t$.
It corresponds to the statistical uncertainty from the
limited number of events used in the fit when using the $W$-boson mass to
constrain the energies of the light quarks from the $W$ decay.

\vspace{0.1in}
\noindent {\bf 2.~~Jet modeling}\\*
\label{DetMod}
\indent The uncertainty in jet modeling has two components for
{\dzero}. This uncertainty is negligible for CDF.

\noindent(i) The jet energy
resolution is smeared for MC jets to match the resolution observed in data,
and the uncertainty on the smearing functions is propagated to $m_t$.

\noindent(ii) The identification
efficiency in MC events is corrected to match that found in data, and the
uncertainty on the correction functions is propagated to $m_t$.

\vspace{0.1in}
\noindent {\bf 3.~~Lepton modeling}\\*
\label{LepPt}
\indent This uncertainty has two components:

\noindent(i) The electron and muon $p_T$ scales are
calibrated to the $J/\psi$ and $Z$-boson mass by both CDF and {\dzero}.
This uncertainty on the calibration is included in the measurements of $m_t$.

\noindent(ii) {\dzero} smears the muon momentum resolution in MC
events to match that in data, and the uncertainty on this correction is
included in this term.
The uncertainty on the electron resolution has a negligible impact on the
measurements of $m_t$.

\vspace{0.15in}
\noindent {\bf 4.~~Signal modeling}\\*
\label{Signal}
\indent There are six components to this uncertainty.
They are combined into one term because the correlations between
channels are similar for each component:

\noindent(i) Knowledge of the PDF parametrization.

\noindent(ii) The quark annihilation and gluon fusion fractions
that differ significantly between leading-log and
next-to-leading-order (NLO) QCD calculations (Run~II).

\noindent(iii) The amount of initial- and final-state radiation
in MC signal events differs from
that in data and is adjusted through the value of $\Lambda_{\rm QCD}$
used in the shower and the scales of time and spacelike showers.

\noindent(iv) Higher-order QCD corrections to
initial- and final-state radiation differ from precise
parton-level models, and this is not accounted for by the choice of scale for
the calculations (Run~II).

\noindent(v) Our model for jet hadronization is based on angular ordering in  {\sc pythia}  
with Tune A underlying-event tuning. Parton showering and 
the underlying event can also be simulated with {\sc herwig} and {\sc jimmy}~\cite{jimmy,tuneA-jimmy}.  
The effect of the  difference on $m_t$
between the two models is included in this term.

\noindent(vi) Final-state partons and remnants of the protons and antiprotons are connected through color strings,
which affect the distributions of jets. Since this effect is
not included in the model for the \ttbar\ signal, the value of $m_t$ has an
uncertainty from this omission (Run~II).

\vspace{0.1in}
\noindent {\bf 5.~~Multiple interactions model}\\*
\label{MHI}
\indent The number of soft $p\bar{p}$ events overlaid on each MC
event has a Poisson distribution. The mean number does not equal
exactly the number seen in data since the luminosity increased as the Tevatron
run progressed. The top-quark mass is measured as a function of the
number of multiple interactions in signal events by CDF,  the signal
MC events are reweighted to match the distribution seen in data
by {\dzero}, and the related uncertainties are included here.


\begingroup
\squeezetable
\begin{table*}[htbp]
\begin{minipage}{5.5in}
\caption[syst-corr]{Correlations in systematic uncertainties (in percent) among the different
measurements of $m_t$.}
\label{systematics-correlations}
\vspace{0.05in}
\begin{ruledtabular}
\begin{tabular}{lll m{0.7cm}  p{0.7cm}   p{0.7cm}   p{0.7cm}  p{0.7cm}  p{0.7cm}   p{0.7cm}  p{0.7cm}  p{0.7cm}  p{0.7cm}  p{0.7cm}   p{0.7cm}}
& &
& \multicolumn{1}{c}{\begin{sideways}CDF~~\end{sideways}}
& \multicolumn{1}{c}{\begin{sideways}{\dzero}\end{sideways}}
& \multicolumn{1}{c}{\begin{sideways}CDF\end{sideways}}
& \multicolumn{1}{c}{\begin{sideways}{\dzero}\end{sideways}}
& \multicolumn{1}{c}{\begin{sideways}CDF\end{sideways}}
& \multicolumn{1}{c}{\begin{sideways}CDF\end{sideways}}
& \multicolumn{1}{c}{\begin{sideways}CDF\end{sideways}}
& \multicolumn{1}{c}{\begin{sideways}{\dzero}\end{sideways}}
& \multicolumn{1}{c}{\begin{sideways}CDF\end{sideways}}
& \multicolumn{1}{c}{\begin{sideways}{\dzero}\end{sideways}}
& \multicolumn{1}{c}{\begin{sideways}CDF\end{sideways}}
& \multicolumn{1}{c}{\begin{sideways}CDF\end{sideways}}
\vspace{0.02in} \\
& &
& \multicolumn{1}{c}{\begin{sideways}Run~II~\end{sideways}}
& \multicolumn{1}{c}{\begin{sideways}Run~II\end{sideways}}
& \multicolumn{1}{c}{\begin{sideways}Run~I\end{sideways}}
& \multicolumn{1}{c}{\begin{sideways}Run~I\end{sideways}}
& \multicolumn{1}{c}{\begin{sideways}Run~II\end{sideways}}
& \multicolumn{1}{c}{\begin{sideways}Run~I\end{sideways}}
& \multicolumn{1}{c}{\begin{sideways}Run~II\end{sideways}}
& \multicolumn{1}{c}{\begin{sideways}Run~II\end{sideways}}
& \multicolumn{1}{c}{\begin{sideways}Run~I\end{sideways}}
& \multicolumn{1}{c}{\begin{sideways}Run~I\end{sideways}}
& \multicolumn{1}{c}{\begin{sideways}Run~II\end{sideways}}
& \multicolumn{1}{c}{\begin{sideways}Run~II\end{sideways}}
\vspace{0.02in} \\
& &
& \multicolumn{1}{c}{\begin{sideways}Lepton+jets\end{sideways}}
& \multicolumn{1}{c}{\begin{sideways}Lepton+jets\end{sideways}}
& \multicolumn{1}{c}{\begin{sideways}Lepton+jets\end{sideways}}
& \multicolumn{1}{c}{\begin{sideways}Lepton+jets\end{sideways}}
& \multicolumn{1}{c}{\begin{sideways}Alljets\end{sideways}}
& \multicolumn{1}{c}{\begin{sideways}Alljets\end{sideways}}
& \multicolumn{1}{c}{\begin{sideways}Dileptons\end{sideways}}
& \multicolumn{1}{c}{\begin{sideways}Dileptons\end{sideways}}
& \multicolumn{1}{c}{\begin{sideways}Dileptons\end{sideways}}
& \multicolumn{1}{c}{\begin{sideways}Dileptons\end{sideways}}
& \multicolumn{1}{c}{\begin{sideways}{\met}+jets\end{sideways}}
& \multicolumn{1}{c}{\begin{sideways}Decay length\end{sideways}} \\
[1.0ex] \hline \\ [-1.5ex]
             &         &                  &   \multicolumn{12}{c}
{\underline{Calibration method}~~~~~\underline{Statistical uncertainty}}
\vspace{0.05in} \\
             &         &                  &   \multicolumn{12}{c}{Not correlated among any measurements}
\vspace{0.1in} \\
             &         &                  &   \multicolumn{12}{c}
{\underline{{\it In situ} light-jet calibration (JES)}}
\vspace{0.05in} \\
Lepton+jets  & Run~II  & ~CDF~~~~~~       &   100 &   ~~~\zp &   ~~~\zp &   ~~~\zp &   ~~~\zp &   ~~~\zp &   ~~~\zp &   ~~~\zp &   ~~~\zp &   ~~~\zp &   ~~~\zp &   ~~~\zp \\
Lepton+jets  & Run~II  & ~{\dzero}~~~~~~  &   ~~~\zp &   100 &   ~~~\zp &   ~~~\zp &   ~~~\zp &   ~~~\zp &   ~~~\zp &   100 &   ~~~\zp &   ~~~\zp &   ~~~\zp &   ~~~\zp \\
Lepton+jets  & Run~I   & ~CDF~~~~~~       &   ~~~\zp &   ~~~\zp &   100 &   ~~~\zp &   ~~~\zp &   ~~~\zp &   ~~~\zp &   ~~~\zp &   ~~~\zp &   ~~~\zp &   ~~~\zp &   ~~~\zp \\
Lepton+jets  & Run~I   & ~{\dzero}~~~~~~  &   ~~~\zp &   ~~~\zp &   ~~~\zp &   100 &   ~~~\zp &   ~~~\zp &   ~~~\zp &   ~~~\zp &   ~~~\zp &   ~~~\zp &   ~~~\zp &   ~~~\zp \\
Alljets      & Run~II  & ~CDF~~~~~~       &   ~~~\zp &   ~~~\zp &   ~~~\zp &   ~~~\zp &   100 &   ~~~\zp &   ~~~\zp &   ~~~\zp &   ~~~\zp &   ~~~\zp &   ~~~\zp &   ~~~\zp \\
Alljets      & Run~I   & ~CDF~~~~~~       &   ~~~\zp &   ~~~\zp &   ~~~\zp &   ~~~\zp &   ~~~\zp &   100 &   ~~~\zp &   ~~~\zp &   ~~~\zp &   ~~~\zp &   ~~~\zp &   ~~~\zp \\
Dileptons    & Run~II  & ~CDF~~~~~~       &   ~~~\zp &   ~~~\zp &   ~~~\zp &   ~~~\zp &   ~~~\zp &   ~~~\zp &   100 &   ~~~\zp &   ~~~\zp &   ~~~\zp &   ~~~\zp &   ~~~\zp \\
Dileptons    & Run~II  & ~{\dzero}~~~~~~  &   ~~~\zp &   100 &   ~~~\zp &   ~~~\zp &   ~~~\zp &   ~~~\zp &   ~~~\zp &   100 &   ~~~\zp &   ~~~\zp &   ~~~\zp &   ~~~\zp \\
Dileptons    & Run~I   & ~CDF~~~~~~       &   ~~~\zp &   ~~~\zp &   ~~~\zp &   ~~~\zp &   ~~~\zp &   ~~~\zp &   ~~~\zp &   ~~~\zp &   100 &   ~~~\zp &   ~~~\zp &   ~~~\zp \\
Dileptons    & Run~I   & ~{\dzero}~~~~~~  &   ~~~\zp &   ~~~\zp &   ~~~\zp &   ~~~\zp &   ~~~\zp &   ~~~\zp &   ~~~\zp &   ~~~\zp &   ~~~\zp &   100 &   ~~~\zp &   ~~~\zp \\
{\met}+jets  & Run~II  & ~CDF~~~~~~       &   ~~~\zp &   ~~~\zp &   ~~~\zp &   ~~~\zp &   ~~~\zp &   ~~~\zp &   ~~~\zp &   ~~~\zp &   ~~~\zp &   ~~~\zp &   100 &   ~~~\zp \\
Decay length & Run~II  & ~CDF~~~~~~       &   ~~~\zp &   ~~~\zp &   ~~~\zp &   ~~~\zp &   ~~~\zp &   ~~~\zp &   ~~~\zp &   ~~~\zp &   ~~~\zp &   ~~~\zp &   ~~~\zp &   100 \\  
\vspace{0.1in} \\
             &         &                  &   \multicolumn{12}{c}
{\underline{Background based on data}}
\vspace{0.05in} \\
Lepton+jets  & Run~II  & ~CDF~~~~~~       &   100 &   ~~~\zp &   ~~~\zp &   ~~~\zp &   ~~~\zp &   ~~~\zp &   ~~~\zp &   ~~~\zp &   ~~~\zp &   ~~~\zp &   ~~~\zp &  100 \\
Lepton+jets  & Run~II  & ~{\dzero}~~~~~~  &   ~~~\zp &   100 &   ~~~\zp &   ~~~\zp &   ~~~\zp &   ~~~\zp &   ~~~\zp &   ~~~\zp &   ~~~\zp &   ~~~\zp &   ~~~\zp &  ~~~\zp \\
Lepton+jets  & Run~I   & ~CDF~~~~~~       &   ~~~\zp &   ~~~\zp &   100 &   ~~~\zp &   ~~~\zp &   ~~~\zp &   ~~~\zp &   ~~~\zp &   ~~~\zp &   ~~~\zp &   ~~~\zp &  ~~~\zp \\
Lepton+jets  & Run~I   & ~{\dzero}~~~~~~  &   ~~~\zp &   ~~~\zp &   ~~~\zp &   100 &   ~~~\zp &   ~~~\zp &   ~~~\zp &   ~~~\zp &   ~~~\zp &   ~~~\zp &   ~~~\zp &  ~~~\zp \\
Alljets      & Run~II  & ~CDF~~~~~~       &   ~~~\zp &   ~~~\zp &   ~~~\zp &   ~~~\zp &   100 &   ~~~\zp &   ~~~\zp &   ~~~\zp &   ~~~\zp &   ~~~\zp &   ~~~\zp &  ~~~\zp \\
Alljets      & Run~I   & ~CDF~~~~~~       &   ~~~\zp &   ~~~\zp &   ~~~\zp &   ~~~\zp &   ~~~\zp &   100 &   ~~~\zp &   ~~~\zp &   ~~~\zp &   ~~~\zp &   ~~~\zp &  ~~~\zp \\
Dileptons    & Run~II  & ~CDF~~~~~~       &   ~~~\zp &   ~~~\zp &   ~~~\zp &   ~~~\zp &   ~~~\zp &   ~~~\zp &   100 &   ~~~\zp &   ~~~\zp &   ~~~\zp &   ~~~\zp &  ~~~\zp \\
Dileptons    & Run~II  & ~{\dzero}~~~~~~  &   ~~~\zp &   ~~~\zp &   ~~~\zp &   ~~~\zp &   ~~~\zp &   ~~~\zp &   ~~~\zp &   100 &   ~~~\zp &   ~~~\zp &   ~~~\zp &  ~~~\zp \\
Dileptons    & Run~I   & ~CDF~~~~~~       &   ~~~\zp &   ~~~\zp &   ~~~\zp &   ~~~\zp &   ~~~\zp &   ~~~\zp &   ~~~\zp &   ~~~\zp &   100 &   ~~~\zp &   ~~~\zp &  ~~~\zp \\
Dileptons    & Run~I   & ~{\dzero}~~~~~~  &   ~~~\zp &   ~~~\zp &   ~~~\zp &   ~~~\zp &   ~~~\zp &   ~~~\zp &   ~~~\zp &   ~~~\zp &   ~~~\zp &   100 &   ~~~\zp &  ~~~\zp \\
{\met}+jets  & Run~II  & ~CDF~~~~~~       &   ~~~\zp &   ~~~\zp &   ~~~\zp &   ~~~\zp &   ~~~\zp &   ~~~\zp &   ~~~\zp &   ~~~\zp &   ~~~\zp &   ~~~\zp &   100 &  ~~~\zp \\
Decay length & Run~II  & ~CDF~~~~~~       &   100 &   ~~~\zp &   ~~~\zp &   ~~~\zp &   ~~~\zp &   ~~~\zp &   ~~~\zp &   ~~~\zp &   ~~~\zp &   ~~~\zp &   ~~~\zp &  100 \\
\vspace{0.1in} \\
             &         &                  &   \multicolumn{12}{c}{\underline{Background from theory}}                                          
\vspace{0.05in} \\
Lepton+jets  & Run~II  & ~CDF~~~~~~       &   100 &   100 &   100 &   100 &   ~~~\zp &   ~~~\zp &   ~~~\zp &   ~~~\zp &   ~~~\zp &   ~~~\zp &   ~~~\zp &   100 \\
Lepton+jets  & Run~II  & ~{\dzero}~~~~~~  &   100 &   100 &   100 &   100 &   ~~~\zp &   ~~~\zp &   ~~~\zp &   ~~~\zp &   ~~~\zp &   ~~~\zp &   ~~~\zp &   100 \\
Lepton+jets  & Run~I   & ~CDF~~~~~~       &   100 &   100 &   100 &   100 &   ~~~\zp &   ~~~\zp &   ~~~\zp &   ~~~\zp &   ~~~\zp &   ~~~\zp &   ~~~\zp &   100 \\
Lepton+jets  & Run~I   & ~{\dzero}~~~~~~  &   100 &   100 &   100 &   100 &   ~~~\zp &   ~~~\zp &   ~~~\zp &   ~~~\zp &   ~~~\zp &   ~~~\zp &   ~~~\zp &   100 \\
Alljets      & Run~II  & ~CDF~~~~~~       &   ~~~\zp &   ~~~\zp &   ~~~\zp &   ~~~\zp &   100 &   100 &   ~~~\zp &   ~~~\zp &   ~~~\zp &   ~~~\zp &   ~~~\zp &   ~~~\zp \\
Alljets      & Run~I   & ~CDF~~~~~~       &   ~~~\zp &   ~~~\zp &   ~~~\zp &   ~~~\zp &   100 &   100 &   ~~~\zp &   ~~~\zp &   ~~~\zp &   ~~~\zp &   ~~~\zp &   ~~~\zp \\
Dileptons    & Run~II  & ~CDF~~~~~~       &   ~~~\zp &   ~~~\zp &   ~~~\zp &   ~~~\zp &   ~~~\zp &   ~~~\zp &   100 &   100 &   100 &   100 &   ~~~\zp &   ~~~\zp \\
Dileptons    & Run~II  & ~{\dzero}~~~~~~  &   ~~~\zp &   ~~~\zp &   ~~~\zp &   ~~~\zp &   ~~~\zp &   ~~~\zp &   100 &   100 &   100 &   100 &   ~~~\zp &   ~~~\zp \\
Dileptons    & Run~I   & ~CDF~~~~~~       &   ~~~\zp &   ~~~\zp &   ~~~\zp &   ~~~\zp &   ~~~\zp &   ~~~\zp &   100 &   100 &   100 &   100 &   ~~~\zp &   ~~~\zp \\
Dileptons    & Run~I   & ~{\dzero}~~~~~~  &   ~~~\zp &   ~~~\zp &   ~~~\zp &   ~~~\zp &   ~~~\zp &   ~~~\zp &   100 &   100 &   100 &   100 &   ~~~\zp &   ~~~\zp \\
{\met}+jets  & Run~II  & ~CDF~~~~~~       &   ~~~\zp &   ~~~\zp &   ~~~\zp &   ~~~\zp &   ~~~\zp &   ~~~\zp &   ~~~\zp &   ~~~\zp &   ~~~\zp &   ~~~\zp &   100 &   ~~~\zp \\
Decay length & Run~II  & ~CDF~~~~~~       &   100 &   100 &   100 &   100 &   ~~~\zp &   ~~~\zp &   ~~~\zp &   ~~~\zp &   ~~~\zp &   ~~~\zp &  ~~~\zp &   100  \\ 
\vspace{0.1in} \\                       
             &         &                  &   \multicolumn{12}{c}
{\underline{Light-jet response (2) (JES)}~~~~~\underline{Offset (JES)}~~~~~\underline{Response to $b/q/g$ jets (JES)}} \\
             &         &                  &   \multicolumn{12}{c}
{\underline{Jet modeling}~~~~~\underline{Lepton modeling}~~~~~\underline{Multiple interactions model}}
\vspace{0.05in} \\
Lepton+jets  & Run~II  & ~CDF~~~~~~       &   100 &   ~~~\zp &   ~~~\zp &   ~~~\zp &   100 &   ~~~\zp &   100 &   ~~~\zp &   ~~~\zp &   ~~~\zp &   100 &   100 \\
Lepton+jets  & Run~II  & ~{\dzero}~~~~~~  &   ~~~\zp &   100 &   ~~~\zp &   ~~~\zp &   ~~~\zp &   ~~~\zp &   ~~~\zp &   100 &   ~~~\zp &   ~~~\zp &   ~~~\zp &   ~~~\zp \\
Lepton+jets  & Run~I   & ~CDF~~~~~~       &   ~~~\zp &   ~~~\zp &   100 &   ~~~\zp &   ~~~\zp &   100 &   ~~~\zp &   ~~~\zp &   100 &   ~~~\zp &   ~~~\zp &   ~~~\zp \\
Lepton+jets  & Run~I   & ~{\dzero}~~~~~~  &   ~~~\zp &   ~~~\zp &   ~~~\zp &   100 &   ~~~\zp &   ~~~\zp &   ~~~\zp &   ~~~\zp &   ~~~\zp &   100 &   ~~~\zp &   ~~~\zp \\
Alljets      & Run~II  & ~CDF~~~~~~       &   100 &   ~~~\zp &   ~~~\zp &   ~~~\zp &   100 &   ~~~\zp &   100 &   ~~~\zp &   ~~~\zp &   ~~~\zp &   100 &   100 \\
Alljets      & Run~I   & ~CDF~~~~~~       &   ~~~\zp &   ~~~\zp &   100 &   ~~~\zp &   ~~~\zp &   100 &   ~~~\zp &   ~~~\zp &   100 &   ~~~\zp &   ~~~\zp &   ~~~\zp \\
Dileptons    & Run~II  & ~CDF~~~~~~       &   100 &   ~~~\zp &   ~~~\zp &   ~~~\zp &   100 &   ~~~\zp &   100 &   ~~~\zp &   ~~~\zp &   ~~~\zp &   100 &   100 \\
Dileptons    & Run~II  & ~{\dzero}~~~~~~  &   ~~~\zp &   100 &   ~~~\zp &   ~~~\zp &   ~~~\zp &   ~~~\zp &   ~~~\zp &   100 &   ~~~\zp &   ~~~\zp &   ~~~\zp &   ~~~\zp \\
Dileptons    & Run~I   & ~CDF~~~~~~       &   ~~~\zp &   ~~~\zp &   100 &   ~~~\zp &   ~~~\zp &   100 &   ~~~\zp &   ~~~\zp &   100 &   ~~~\zp &   ~~~\zp &   ~~~\zp \\
Dileptons    & Run~I   & ~{\dzero}~~~~~~  &   ~~~\zp &   ~~~\zp &   ~~~\zp &   100 &   ~~~\zp &   ~~~\zp &   ~~~\zp &   ~~~\zp &   ~~~\zp &   100 &   ~~~\zp &   ~~~\zp \\
{\met}+jets  & Run~II  & ~CDF~~~~~~       &   100 &   ~~~\zp &   ~~~\zp &   ~~~\zp &   100 &   ~~~\zp &   100 &   ~~~\zp &   ~~~\zp &   ~~~\zp &   100 &   100 \\
Decay length & Run~II  & ~CDF~~~~~~       &   100 &   ~~~\zp &   ~~~\zp &   ~~~\zp &   100 &   ~~~\zp &   100 &   ~~~\zp &   ~~~\zp &   ~~~\zp &   100 &   100 \\  
\vspace{0.05in}
\end{tabular}
\end{ruledtabular}
\end{minipage}
\end{table*}
\endgroup

\begingroup
\squeezetable
\begin{table*}[htbp]
\begin{minipage}{5.5in}
\caption[syst-corr]{Correlations in systematic uncertainties (in percent) among the different
measurements of $m_t$ (continued).}
\label{systematics-correlations2}
\vspace{0.05in}
\begin{ruledtabular}
\begin{tabular}{lll m{0.7cm}  p{0.7cm}   p{0.7cm}   p{0.7cm}  p{0.7cm}  p{0.7cm}   p{0.7cm}  p{0.7cm}  p{0.7cm}  p{0.7cm}  p{0.7cm}   p{0.7cm}}
& &
& \multicolumn{1}{c}{\begin{sideways}CDF~~\end{sideways}}
& \multicolumn{1}{c}{\begin{sideways}{\dzero}\end{sideways}}
& \multicolumn{1}{c}{\begin{sideways}CDF\end{sideways}}
& \multicolumn{1}{c}{\begin{sideways}{\dzero}\end{sideways}}
& \multicolumn{1}{c}{\begin{sideways}CDF\end{sideways}}
& \multicolumn{1}{c}{\begin{sideways}CDF\end{sideways}}
& \multicolumn{1}{c}{\begin{sideways}CDF\end{sideways}}
& \multicolumn{1}{c}{\begin{sideways}{\dzero}\end{sideways}}
& \multicolumn{1}{c}{\begin{sideways}CDF\end{sideways}}
& \multicolumn{1}{c}{\begin{sideways}{\dzero}\end{sideways}}
& \multicolumn{1}{c}{\begin{sideways}CDF\end{sideways}}
& \multicolumn{1}{c}{\begin{sideways}CDF\end{sideways}}
\vspace{0.02in} \\
& &
& \multicolumn{1}{c}{\begin{sideways}Run~II~\end{sideways}}
& \multicolumn{1}{c}{\begin{sideways}Run~II\end{sideways}}
& \multicolumn{1}{c}{\begin{sideways}Run~I\end{sideways}}
& \multicolumn{1}{c}{\begin{sideways}Run~I\end{sideways}}
& \multicolumn{1}{c}{\begin{sideways}Run~II\end{sideways}}
& \multicolumn{1}{c}{\begin{sideways}Run~I\end{sideways}}
& \multicolumn{1}{c}{\begin{sideways}Run~II\end{sideways}}
& \multicolumn{1}{c}{\begin{sideways}Run~II\end{sideways}}
& \multicolumn{1}{c}{\begin{sideways}Run~I\end{sideways}}
& \multicolumn{1}{c}{\begin{sideways}Run~I\end{sideways}}
& \multicolumn{1}{c}{\begin{sideways}Run~II\end{sideways}}
& \multicolumn{1}{c}{\begin{sideways}Run~II\end{sideways}}
\vspace{0.02in} \\
& &
& \multicolumn{1}{c}{\begin{sideways}Lepton+jets\end{sideways}}
& \multicolumn{1}{c}{\begin{sideways}Lepton+jets\end{sideways}}
& \multicolumn{1}{c}{\begin{sideways}Lepton+jets\end{sideways}}
& \multicolumn{1}{c}{\begin{sideways}Lepton+jets\end{sideways}}
& \multicolumn{1}{c}{\begin{sideways}Alljets\end{sideways}}
& \multicolumn{1}{c}{\begin{sideways}Alljets\end{sideways}}
& \multicolumn{1}{c}{\begin{sideways}Dileptons\end{sideways}}
& \multicolumn{1}{c}{\begin{sideways}Dileptons\end{sideways}}
& \multicolumn{1}{c}{\begin{sideways}Dileptons\end{sideways}}
& \multicolumn{1}{c}{\begin{sideways}Dileptons\end{sideways}}
& \multicolumn{1}{c}{\begin{sideways}{\met}+jets\end{sideways}}
& \multicolumn{1}{c}{\begin{sideways}Decay length\end{sideways}} \\
[1.0ex] \hline \\ [-1.5ex]
             &         &                  &   \multicolumn{12}{c}
{\underline{Light-jet response (1) (JES)}}
\vspace{0.05in} \\
Lepton+jets  & Run~II  & ~CDF~~~~~~       &   100 &   ~~~\zp &   100 &   ~~~\zp &   100 &   100 &   100 &   ~~~\zp &   100 &   ~~~\zp &   100 &   100 \\
Lepton+jets  & Run~II  & ~{\dzero}~~~~~~  &   ~~~\zp &   100 &   ~~~\zp &   100 &   ~~~\zp &   ~~~\zp &   ~~~\zp &   100 &   ~~~\zp &   100 &   ~~~\zp &   ~~~\zp \\
Lepton+jets  & Run~I   & ~CDF~~~~~~       &   100 &   ~~~\zp &   100 &   ~~~\zp &   100 &   100 &   100 &   ~~~\zp &   100 &   ~~~\zp &   100 &   100 \\
Lepton+jets  & Run~I   & ~{\dzero}~~~~~~  &   ~~~\zp &   100 &   ~~~\zp &   100 &   ~~~\zp &   ~~~\zp &   ~~~\zp &   100 &   ~~~\zp &   100 &   ~~~\zp &   ~~~\zp \\
Alljets      & Run~II  & ~CDF~~~~~~       &   100 &   ~~~\zp &   100 &   ~~~\zp &   100 &   100 &   100 &   ~~~\zp &   100 &   ~~~\zp &   100 &   100 \\
Alljets      & Run~I   & ~CDF~~~~~~       &   100 &   ~~~\zp &   100 &   ~~~\zp &   100 &   100 &   100 &   ~~~\zp &   100 &   ~~~\zp &   100 &   100 \\
Dileptons    & Run~II  & ~CDF~~~~~~       &   100 &   ~~~\zp &   100 &   ~~~\zp &   100 &   100 &   100 &   ~~~\zp &   100 &   ~~~\zp &   100 &   100 \\
Dileptons    & Run~II  & ~{\dzero}~~~~~~  &   ~~~\zp &   100 &   ~~~\zp &   100 &   ~~~\zp &   ~~~\zp &   ~~~\zp &   100 &   ~~~\zp &   100 &   ~~~\zp &   ~~~\zp \\
Dileptons    & Run~I   & ~CDF~~~~~~       &   100 &   ~~~\zp &   100 &   ~~~\zp &   100 &   100 &   100 &   ~~~\zp &   100 &   ~~~\zp &   100 &   100 \\
Dileptons    & Run~I   & ~{\dzero}~~~~~~  &   ~~~\zp &   100 &   ~~~\zp &   100 &   ~~~\zp &   ~~~\zp &   ~~~\zp &   100 &   ~~~\zp &   100 &   ~~~\zp &   ~~~\zp \\
{\met}+jets  & Run~II  & ~CDF~~~~~~       &   100 &   ~~~\zp &   100 &   ~~~\zp &   100 &   100 &   100 &   ~~~\zp &   100 &   ~~~\zp &   100 &   100 \\
Decay length & Run~II  & ~CDF~~~~~~       &   100 &   ~~~\zp &   100 &   ~~~\zp &   100 &   100 &   100 &   ~~~\zp &   100 &   ~~~\zp &   100 &   100 \\
\vspace{0.1in} \\
             &         &                  &   \multicolumn{12}{c}
{\underline{Out-of-cone correction (JES)}~~~~~\underline{Model for $b$ jets (JES)}~~~~~\underline{Signal modeling}}
\vspace{0.05in} \\
             &         &                  &   \multicolumn{12}{c}{100\% correlated among all measurements}
\vspace{0.05in}
\end{tabular}
\end{ruledtabular}
\end{minipage}
\end{table*}
\endgroup

\vspace{0.15in}
\noindent {\bf 6.~~Background from theory}\\*
\label{BGMC}
\indent There are four components in this uncertainty:

\noindent(i) Difference between NLO calculations of the fraction of
heavy-flavor jets in $W$+jets events.
The {\sc alpgen} model underestimates this fraction.

\noindent(ii) Impact of factorization and renormalization scales on the
$W$+jets simulation, which affects the background model for
distributions characterizing jets.

\noindent(iii) 
The theoretical cross sections 
used to normalize all MC estimated background processes
(except for $W$+jets for CDF and {\dzero} lepton+jets
measurements, and Drell-Yan production for CDF dilepton measurements).

\noindent(iv) Impact of difference between the MC modeling of background
kinematic distributions and those observed in data.

\vspace{0.1in}
\noindent {\bf 7.~~Background based on data}\\*
\label{BGData}
\indent This refers primarily to uncertainties 
from the normalization of certain background components to data. These include
multijet backgrounds in the lepton+jets, alljets, and {\met}+jets
analyses, the $W$+jets background in the {\dzero} lepton+jets analyses,
and the Drell-Yan backgrounds in the CDF dilepton analyses.

{\dzero} also considers the following four components of uncertainty:

\noindent(i) The uncertainty from correcting the MC events to match the
trigger efficiency in data,  which is based on the turn-on response for each
trigger element. 

\noindent(ii) The uncertainty from applying tag-rate and
taggability corrections to MC events to make the
efficiencies match the data for each jet flavor.


\noindent(iii) The uncertainty on the fraction of multijet events included in the pseudoexperiments used for calibration.

\vspace{0.15in}
\noindent {\bf 8.~~Calibration method}\\*
\label{Method}
\indent The extracted values of $m_t$ are calibrated using a straight-line
fit to the relationship between input mass and measured mass in simulated pseudoexperiments. This
term includes the systematic uncertainties from the slope and offset
of this calibration.


\begingroup
\squeezetable
\begin{table*}[!h!tbp]
\caption[meas-corr-weights]{Correlations in \% among the input $m_t$
measurements and their weights in the BLUE combination.}
\label{measurement-correlations-weights}
\vspace{0.05in}
\begin{ruledtabular}
\begin{tabular}{lllrrrrrrrrrrrrc.{2.4}}
& &
& \multicolumn{1}{c}{\begin{sideways}CDF~~\end{sideways}}
& \multicolumn{1}{c}{\begin{sideways}{\dzero}\end{sideways}}
& \multicolumn{1}{c}{\begin{sideways}CDF\end{sideways}}
& \multicolumn{1}{c}{\begin{sideways}{\dzero}\end{sideways}}
& \multicolumn{1}{c}{\begin{sideways}CDF\end{sideways}}
& \multicolumn{1}{c}{\begin{sideways}CDF\end{sideways}}
& \multicolumn{1}{c}{\begin{sideways}CDF\end{sideways}}
& \multicolumn{1}{c}{\begin{sideways}{\dzero}\end{sideways}}
& \multicolumn{1}{c}{\begin{sideways}CDF\end{sideways}}
& \multicolumn{1}{c}{\begin{sideways}{\dzero}\end{sideways}}
& \multicolumn{1}{c}{\begin{sideways}CDF\end{sideways}}
& \multicolumn{1}{c}{\begin{sideways}CDF\end{sideways}}
& & \vspace{0.02in} \\
& &
& \multicolumn{1}{c}{\begin{sideways}Run~II~\end{sideways}}
& \multicolumn{1}{c}{\begin{sideways}Run~II\end{sideways}}
& \multicolumn{1}{c}{\begin{sideways}Run~I\end{sideways}}
& \multicolumn{1}{c}{\begin{sideways}Run~I\end{sideways}}
& \multicolumn{1}{c}{\begin{sideways}Run~II\end{sideways}}
& \multicolumn{1}{c}{\begin{sideways}Run~I\end{sideways}}
& \multicolumn{1}{c}{\begin{sideways}Run~II\end{sideways}}
& \multicolumn{1}{c}{\begin{sideways}Run~II\end{sideways}}
& \multicolumn{1}{c}{\begin{sideways}Run~I\end{sideways}}
& \multicolumn{1}{c}{\begin{sideways}Run~I\end{sideways}}
& \multicolumn{1}{c}{\begin{sideways}Run~II\end{sideways}}
& \multicolumn{1}{c}{\begin{sideways}Run~II\end{sideways}}
& & \vspace{0.02in} \\
& &
& \multicolumn{1}{c}{\begin{sideways}Lepton+jets\end{sideways}}
& \multicolumn{1}{c}{\begin{sideways}Lepton+jets\end{sideways}}
& \multicolumn{1}{c}{\begin{sideways}Lepton+jets\end{sideways}}
& \multicolumn{1}{c}{\begin{sideways}Lepton+jets\end{sideways}}
& \multicolumn{1}{c}{\begin{sideways}Alljets\end{sideways}}
& \multicolumn{1}{c}{\begin{sideways}Alljets\end{sideways}}
& \multicolumn{1}{c}{\begin{sideways}Dileptons\end{sideways}}
& \multicolumn{1}{c}{\begin{sideways}Dileptons\end{sideways}}
& \multicolumn{1}{c}{\begin{sideways}Dileptons\end{sideways}}
& \multicolumn{1}{c}{\begin{sideways}Dileptons\end{sideways}}
& \multicolumn{1}{c}{\begin{sideways}{\met}+jets\end{sideways}}
& \multicolumn{1}{c}{\begin{sideways}Decay length\end{sideways}}
& & \multicolumn{1}{c}{Weight~~~~~~~~~~~~~} \\
[1.0ex] \hline \\ [-1.5ex]
Lepton+jets  & Run~II  & ~CDF~~~~~~~~~~~~ & 100 &  27 &   45 &   25 &   25 &   26 &   44 &   12 &   26 &   11 &   24 &    8 & ~~~~~~ & $55.50$ \\
Lepton+jets  & Run~II  & ~{\dzero}        &  27 & 100 &   21 &   14 &   16 &    9 &   11 &   39 &   13 &    7 &   15 &    6 &        & $26.66$ \\
Lepton+jets  & Run~I   & ~CDF             &  45 &  21 &  100 &   26 &   25 &   32 &   54 &   12 &   29 &   11 &   22 &    7 &        & $-4.72$ \\
Lepton+jets  & Run~I   & ~{\dzero}        &  25 &  14 &   26 &  100 &   12 &   14 &   27 &    7 &   15 &   16 &   10 &    5 &        & $-0.06$ \\
Alljets      & Run~II  & ~CDF             &  25 &  16 &   25 &   12 &  100 &   15 &   25 &   10 &   15 &    7 &   14 &    4 &        & $13.99$ \\
Alljets      & Run~I   & ~CDF             &  26 &   9 &   32 &   14 &   15 &  100 &   38 &    6 &   19 &    7 &   14 &    4 &        & $-0.80$ \\
Dileptons    & Run~II  & ~CDF             &  44 &  11 &   54 &   27 &   25 &   38 &  100 &    7 &   32 &   13 &   22 &    6 &        &  $1.41$ \\
Dileptons    & Run~II  & ~{\dzero}        &  12 &  39 &   12 &    7 &   10 &    6 &    7 &  100 &    8 &    5 &   10 &    3 &        &  $2.28$ \\
Dileptons    & Run~I   & ~CDF             &  26 &  13 &   29 &   15 &   15 &   19 &   32 &    8 &  100 &    8 &   14 &    4 &        & $-1.05$ \\
Dileptons    & Run~I   & ~{\dzero}        &  11 &   7 &   11 &   16 &    7 &    7 &   13 &    5 &    8 &  100 &    6 &    2 &        & $-0.15$ \\
{\met}+jets  & Run~II  & ~CDF             &  24 &  15 &   22 &   10 &   14 &   14 &   22 &   10 &   14 &    6 &  100 &    4 &        &  $6.65$ \\
Decay length & Run~II  & ~CDF             &   8 &   6 &    7 &    5 &    4 &    4 &    6 &    3 &    4 &    2 &    4 &  100 &        &  $0.29$   
\vspace{0.02in} \\
\end{tabular}
\end{ruledtabular}
\end{table*}
\endgroup


\vspace{0.25in}
\noindent {\bf 9.~~Statistical uncertainty}\\*
\label{Statistics}
\indent The statistical uncertainties are determined from the number of data events
in each of the 12 measurements.

\begin{figure}[!h!tb]
\includegraphics[width=3.5in,height=2.in]{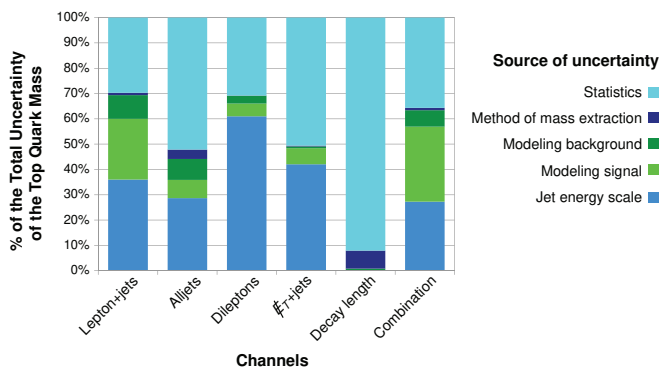}
\vspace{-0.2in}
\caption[uncertainties]{The average uncertainties for CDF and {\dzero}
for each Run~II measurement and for the Tevatron
combination, separated according to major components. (See Table~\ref{ljets-uncertainties} in the Appendix
for details on the systematic categories. In this figure, the jet and lepton modeling systematic uncertainties
are grouped into the modeling background category.)}
\label{uncertainty-percent}
\end{figure}

\vspace{0.2in}
Figure~\ref{uncertainty-percent} shows the relative contribution for each major
uncertainty to the analysis channels in Run~II. The Appendix
provides more detail on how each of the sources of
the uncertainties is estimated.

\vspace{-0.15in}
\subsection{Uncertainty correlations}
\label{uncertainty-correlations}

Tables~\ref{systematics-correlations} and \ref{systematics-correlations2} indicate how uncertainties are
correlated between measurements. There are seven patterns
of correlation:

\vspace{0.1in}
\noindent(i)~Statistical uncertainty and calibration method uncertainty are not correlated among the measurements.

\noindent(ii)~Correlations among {\dzero} measurements that implement the 
same final jet energy corrections for the uncertainty from {\it in situ} light-jet calibration.

\noindent(iii)~Correlations among CDF measurements that use
the same data samples for the uncertainty from background based on data.

\noindent(iv)~Correlations among all measurements in the same
$t\bar{t}$ decay channel for the uncertainty from background estimated from theory.

\noindent(v)~Correlations of measurements within the same experiment
for a given run period for the uncertainties from
light-jet response (2), 
offset, response to $b/q/g$ jets, 
jet modeling, 
lepton modeling and
multiple interactions model.

\noindent(vi)~Correlations for measurements within the same experiment
such as the uncertainty from light-jet response (1).

\noindent(vii)~Correlations among all measurements such as the uncertainties from
out-of-cone correction, model for $b$ jets,
and signal modeling.

We assume that all sources correspond to either no or 100\% correlation. A check of
this assumption (see Sec.~\ref{consistency-checks}) shows that it has
a negligible effect on the combined value and uncertainty of $m_t$.


\begin{figure*}[!h!tb]
\includegraphics[width=6.5in]{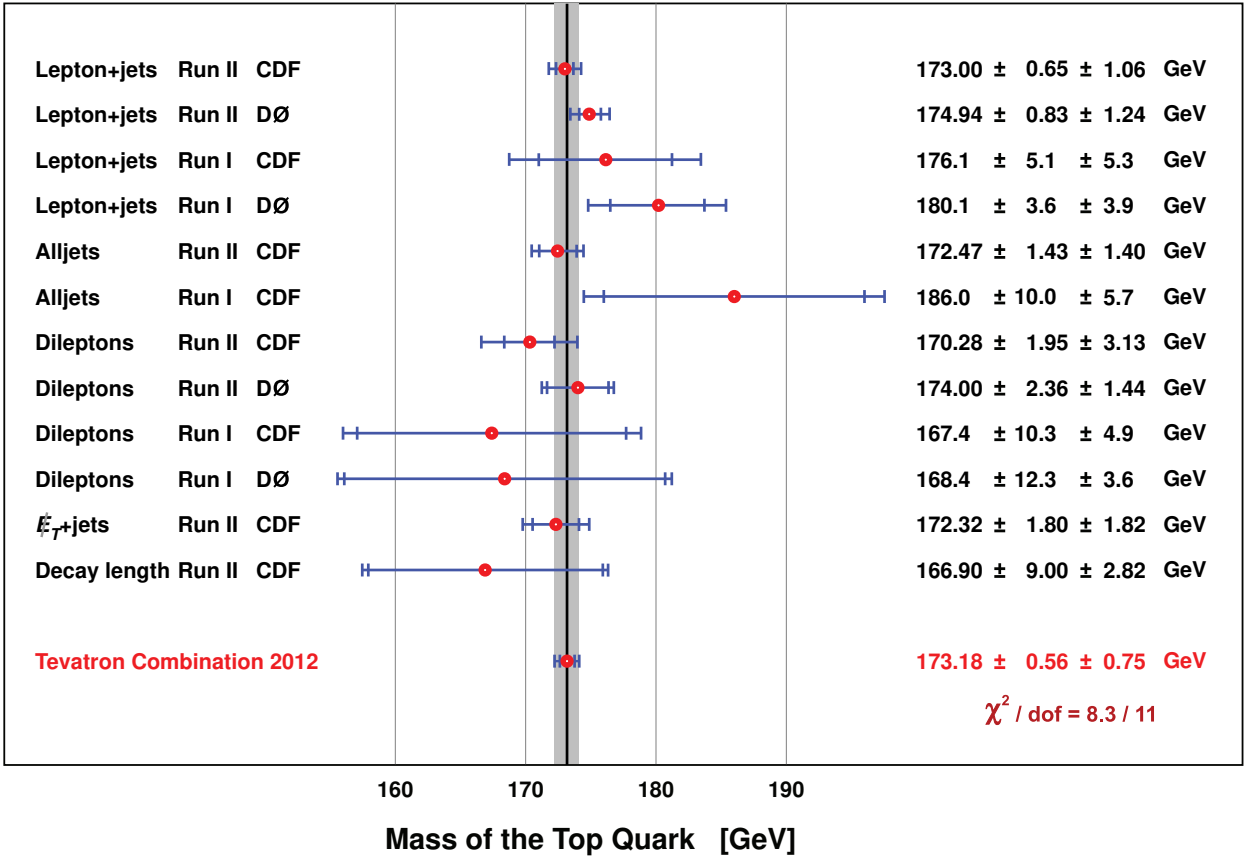}
\caption[top-mass]{The 12 input measurements of $m_t$ from the Tevatron
collider experiments along with the resulting combined value of $m_t^{\rm comb}$.
The gray region corresponds to $\pm 0.94$~GeV.}
\label{top-mass-plot}
\end{figure*}


\vspace{-0.15in}
\subsection{Measurement correlations}
\label{measurement-correlations}
\vspace{-0.1in}

The uncertainties shown in Table~\ref{uncertainties} and their
correlations shown in Tables~\ref{systematics-correlations} and \ref{systematics-correlations2} provide the
correlations among the 12 input values of $m_t$.
The correlation matrix for these measurements, as returned by the combination procedure, is
shown in Table~\ref{measurement-correlations-weights}. The inversion
of the covariance matrix built with the correlation matrix defines the measurement weights, as described
in Sec.~\ref{BLUE-method}. 


\vspace{-0.15in}
\subsection{Measurement weights}
\label{measurement-weights}
\vspace{-0.1in}

As discussed in Sec.~\ref{BLUE-method}, the combined mass $m_t^{\rm comb}$ 
is defined through the set of
weights that minimize the squared difference between $m_t^{\rm comb}$ and the true value of $m_t$, 
which is equivalent to minimizing the sum of the covariance matrix elements.
Table~\ref{measurement-correlations-weights} gives the weights $w_i$
for each of the input measurements as determined in this
minimization. A weight of zero means that an input measurement has no
effect on $m_t^{\rm comb}$. The Run~I measurement weights are negative, which
reflects the fact that the correlations for these and other measurements are larger
than the ratio of their total uncertainties~\cite{BLUE-method-2}. In
this case, the less precise measurement may acquire a negative
weight. Input measurements with negative weights still affect the value 
of $m_t^{\rm comb}$ and reduce the total uncertainty.
By design, the sum of the weights is set to unity.

%
%
%

\begingroup
\squeezetable
\begin{table*}[!h!tb]
\begin{minipage}{5.5in}
\caption[separate-results]{Separate calculations of $m_t^{\rm comb}$ for each
$t\bar{t}$ decay mode, by run period, and by experiment, and
their $\chi^2$ probabilities.}
\label{separate-results}
\vspace{0.05in}
\begin{ruledtabular}
\begin{tabular}{lccccccccccccccccc}
Subset & $m_t^{\rm comb}$  & & \multicolumn{6}{c}{Consistency $\chi^2$}     & & \multicolumn{6}{c}{  $\chi^2$ probability } \\
       &                   & & \multicolumn{6}{c}{(Degrees of freedom = 1)} & & \multicolumn{6}{c}{ }                         \\
 & & & 
\multicolumn{1}{c}{\begin{sideways}Lepton+jets~~\end{sideways}} &
\multicolumn{1}{c}{\begin{sideways}Alljets~~\end{sideways}}     &
\multicolumn{1}{c}{\begin{sideways}Dileptons~~\end{sideways}}   &
\multicolumn{1}{c}{\begin{sideways}{\met}+jets~~\end{sideways}} &
\multicolumn{1}{c}{\begin{sideways}Run~II -- Run~I~~\end{sideways}} &
\multicolumn{1}{c}{\begin{sideways}CDF -- {\dzero}~~\end{sideways}} & &
\multicolumn{1}{c}{\begin{sideways}Lepton+jets~~\end{sideways}} &
\multicolumn{1}{c}{\begin{sideways}Alljets~~\end{sideways}}     &
\multicolumn{1}{c}{\begin{sideways}Dileptons~~\end{sideways}}   &
\multicolumn{1}{c}{\begin{sideways}{\met}+jets~~\end{sideways}} &
\multicolumn{1}{c}{\begin{sideways}Run~II -- Run~I~~\end{sideways}} &
\multicolumn{1}{c}{\begin{sideways}CDF -- {\dzero}~~\end{sideways}} \\
[0.5ex] \hline \\ [-1.5ex]
Lepton+jets & $173.4 \pm 1.0$ & ~~ & ---  & 0.14 & 1.51 & 0.28 &      &                       & ~~ & ---  & 71\% & 22\% & 60\% &      &                       \\
Alljets     & $172.7 \pm 1.9$ & ~~ & 0.14 & ---  & 0.40 & 0.04 &      &                       & ~~ & 71\% & ---  & 53\% & 85\% &      &                       \\
Dileptons   & $171.1 \pm 2.1$ & ~~ & 1.51 & 0.40 & ---  & 0.12 &      &                       & ~~ & 22\% & 53\% & ---  & 73\% &      &                       \\
{\met}+jets & $172.1 \pm 2.5$ & ~~ & 0.28 & 0.04 & 0.12 & ---  &      &                       & ~~ & 60\% & 85\% & 73\% & ---  &      &                       \\
Run II      & $173.6 \pm 1.0$ & ~~ &      &      &      &      & \multirow{2}{*}{2.89} &      & ~~ &      &      &      &      & \multirow{2}{*}{9\%} &       \\
Run I       & $180.0 \pm 4.1$ & ~~ &      &      &      &      &      &                       & ~~ &      &      &      &      &      &                       \\
CDF         & $172.5 \pm 1.0$ & ~~ &      &      &      &      &      & \multirow{2}{*}{2.56} & ~~ &      &      &      &      &      & \multirow{2}{*}{11\%} \\
{\dzero}    & $174.9 \pm 1.4$ & ~~ &      &      &      &      &      &                       & ~~ &      &      &      &      &      &                          
\vspace{0.02in}
\end{tabular}
\end{ruledtabular}
\end{minipage}
\end{table*}
\endgroup


\section{Results of the combination}
\label{results-combination}

\subsection{Tevatron top-quark mass result}
\label{mass-result}

Combining the 12 independent
measurements of $m_t$ from the CDF and {\dzero} Collaborations yields
\begin{align*}
m_t^{\rm comb} & = 173.18 \pm 0.56 \left({\rm stat}\right)
                          \pm 0.75 \left({\rm syst}\right){\rm ~GeV}
\\
               & = 173.18 \pm 0.94{\rm ~GeV}.
\end{align*}
The uncertainties are split into their components in
Table~\ref{uncertainties} and Fig.~\ref{uncertainty-percent}. The
jet energy scale contributes 0.49~GeV to the total systematic
uncertainty. Of this, 0.39~GeV arises from limited statistics
of the {\it in situ} JES calibration and 0.30~GeV from the 
remaining contributions.
Figure~\ref{top-mass-plot} summarizes the input $m_t$ values
and the combined result. 

We assess the consistency of the input $m_t$ measurements
with their combination using a $\chi^2$ test statistic, defined as
follows:
\begin{multline*}
\chi^2_{\rm comb} =
   \left(\boldsymbol{m_t^i}
   - \boldsymbol{m_t^{\rm comb}}\right)^T \\
   \times {\rm Covariance}^{-1} \! \left(m_t^i,m_t^j\right)
   \left(\boldsymbol{m_t^j} - \boldsymbol{m_t^{\rm comb}}\right),
\end{multline*}
where $\boldsymbol{m_t^i}$ is a column vector of the 12 $m_t$
inputs, $\boldsymbol{m_t^{\rm comb}}$ is a
matching column vector for the measurements adjusted in the previous minimization, and the
superscript $T$ denotes the transpose. We find
\begin{align*}
\chi^2_{\rm comb} = 8.3{\rm ~for~11~degrees~of~freedom},
\end{align*}
which is equivalent to a 69\% probability for agreement (i.e., $p$ value for the observed $\chi^2$ value) 
among the 12  input measurements.


\subsection{Consistency checks}
\label{consistency-checks}

We check one aspect of the assumption that biases in the input
$m_t$ are on average zero (see Sec.~\ref{BLUE-method}) by calculating separately the combined $m_t^{\rm comb}$
for each $t\bar{t}$ decay mode, each run period, and each experiment.
The results are shown in Table~\ref{separate-results}. The
resulting $m_t^{\rm comb}$ values are calculated using all 12 input measurements
and their correlations. The $\chi^2$ test statistic provides
the compatibility of each subset with the others and is defined as
\begin{multline*}
\chi^2_{\rm sub1,sub2} = \\
    \left(m_t^{\rm sub1} - m_t^{\rm sub2}\right)^2
    {\rm Covariance}^{-1} \! \left(m_t^{\rm sub1},m_t^{\rm sub2}\right).
\end{multline*}
The $\chi^2$ values in Table~\ref{separate-results} show that
biases in the input measurements are not large.

To check the impact of the assumption
that the systematic uncertainty terms are either 0\% or 100\%
correlated between input measurements, we change all off-diagonal
100\% values to 50\% (see Tables~\ref{systematics-correlations} and \ref{systematics-correlations2}) and
recalculate the combined top-quark mass. This extreme change shifts the
central mass value up by 0.17~GeV and reduces the uncertainty negligibly.
The chosen approach is therefore conservative.


\subsection{Summary}
\label{summary}

We have combined 12 measurements of the mass of the top quark by the CDF and
{\dzero} collaborations at the Tevatron collider and find
\begin{align*}
m_t^{\rm comb} & = 173.18 \pm 0.56 \left({\rm stat}\right)
                          \pm 0.75 \left({\rm syst}\right){\rm ~GeV}
\end{align*}
which corresponds to a precision of 0.54\%.
The result is shown in Table~\ref{mass-summary} together with previous
combined results for comparison. The input measurements for this
combination use up to 5.8~fb$^{-1}$ of integrated luminosity for each experiment, while 10~fb$^{-1}$ are
now available. We therefore expect the final 
combination to improve in precision with the use of all the data, but also from
analyzing all $t\bar{t}$ decay channels in both experiments, and from
the application of improved measurement techniques, signal and
background models, and calibration corrections to all channels that will
reduce systematic uncertainties. 
Currently, there are also some overlaps of the systematic effects that are included in different uncertainty categories.
In addition to the {\em in situ} light-jet calibration systematic uncertainty that will scale down with the increase of analyzed luminosity, 
these levels of double counting are expected to be reduced for the next combination.
The combination presented here
has a 0.54\% precision on $m_t$, making the top quark 
the particle with the best known mass in the SM.

\begingroup
\squeezetable
\begin{table}[!h!tb]
\caption[mass-summary]{Mass measurements of the top quark from 1999 until this publication at the Tevatron collider.}
\label{mass-summary}
\vspace{0.05in}
\begin{ruledtabular}
\begin{tabular}{l.{1.1}.{3.2}c.{2.2}c.{1.2}.{1.2}c}
\multicolumn{9}{c}{   } \vspace{-0.07in} \\
Year         & \multicolumn{1}{c}{Integrated}  & \multicolumn{5}{c}{$m_t$} & \multicolumn{1}{c}{Uncertainty} & Reference \\
             & \multicolumn{1}{c}{luminosity}  & \multicolumn{5}{c}{[GeV]}      & \multicolumn{1}{c}{on}          &           \\
             & \multicolumn{1}{c}{[fb$^{-1}$]} & \multicolumn{5}{c}{ }     & \multicolumn{1}{c}{$m_t$}        &           \\
[1.0ex] \hline \\ [-1.5ex]
1999  &  0.1  &  174.3  & $\!\!\pm\!\!\!$ &  3.2 & $\!\!\pm\!\!$ &  4.0  &  2.9\% & \cite{tevatron-combination-1}  \\
2004  &  0.1  &  178.0  & $\!\!\pm\!\!\!$ &  2.7 & $\!\!\pm\!\!$ &  3.3  &  2.4\% & \cite{tevatron-combination-2}  \\
2005  &  0.3  &  172.7  & $\!\!\pm\!\!\!$ &  1.7 & $\!\!\pm\!\!$ &  2.4  &  1.7\% & \cite{tevatron-combination-3}  \\
2006  &  0.7  &  172.5  & $\!\!\pm\!\!\!$ &  1.3 & $\!\!\pm\!\!$ &  1.9  &  1.3\% & \cite{tevatron-combination-4}  \\
2006  &  1.0  &  171.4  & $\!\!\pm\!\!\!$ &  1.2 & $\!\!\pm\!\!$ &  1.8  &  1.2\% & \cite{tevatron-combination-5}  \\
2007  &  2.1  &  170.9  & $\!\!\pm\!\!\!$ &  1.1 & $\!\!\pm\!\!$ &  1.5  &  1.1\% & \cite{tevatron-combination-6}  \\
2008  &  2.1  &  172.6  & $\!\!\pm\!\!\!$ &  0.8 & $\!\!\pm\!\!$ &  1.1  &  0.8\% & \cite{tevatron-combination-7}  \\
2008  &  2.1  &  172.4  & $\!\!\pm\!\!\!$ &  0.7 & $\!\!\pm\!\!$ &  1.0  &  0.7\% & \cite{tevatron-combination-8}  \\
2009  &  3.6  &  173.1  & $\!\!\pm\!\!\!$ &  0.6 & $\!\!\pm\!\!$ &  1.1  &  0.7\% & \cite{tevatron-combination-9}  \\
2010  &  5.6  &  173.32 & $\!\!\pm\!\!\!$ & 0.56 & $\!\!\pm\!\!$ & 0.89  & 0.61\% & \cite{tevatron-combination-10} \\
2011  &  5.8  &  173.18 & $\!\!\pm\!\!\!$ & 0.56 & $\!\!\pm\!\!$ & 0.75  & 0.54\% & \cite{tevatron-combination-11} \\  
[1.0ex] \hline \\ [-1.5ex]
      &  5.8  &  173.18 & $\!\!\pm\!\!\!$ & 0.56 & $\!\!\pm\!\!$ & 0.75  & 0.54\% & This paper    
\vspace{0.02in}
\end{tabular}
\end{ruledtabular}
\end{table}
\endgroup


\section*{\acknowledgmentsname}
\label{acknowledgments}
\vspace{-0.1in}
We thank the Fermilab staff and technical staffs of the participating institutions for their vital contributions and acknowledge support from the
DOE and NSF (USA),
ARC (Australia),
CNPq, FAPERJ, FAPESP and FUNDUNESP (Brazil),
NSERC (Canada),
NSC, CAS and CNSF (China),
Colciencias (Colombia),
MSMT and GACR (Czech Republic),
the Academy of Finland,
CEA and CNRS/IN2P3 (France),
BMBF and DFG (Germany),
DAE and DST (India),
SFI (Ireland),
INFN (Italy),
MEXT (Japan),
the Korean World Class University Program and NRF (Korea),
CONACyT (Mexico),
FOM (Netherlands),
MON, NRC KI and RFBR (Russia),
the Slovak R\&D Agency, 
the Ministerio de Ciencia e Innovaci\'{o}n, and Programa Consolider-Ingenio 2010 (Spain),
The Swedish Research Council (Sweden),
SNSF (Switzerland),
STFC and the Royal Society (United Kingdom),
and the A.P Sloan Foundation (USA).


%
%
%

\appendix
\section*{APPENDIX:~~EVALUATION OF SYSTEMATIC UNCERTAINTIES}
\label{appendix-uncertainties}


Systematic uncertainties arise from inadequate modeling of signal
and backgrounds and from the inability to reproduce the detector response with 
simulated events. Systematic uncertainties also arise from ambiguities 
in reconstructing the top quarks from their jet  
and lepton remnants. We minimize such uncertainties by using 
independent data to calibrate the absolute response of the detector, 
and we use state-of-the-art input from theory for modeling the 
signal and backgrounds.   
We use alternative models for signal and different 
parameters for modeling backgrounds to check our assumptions. 

Table~\ref{ljets-uncertainties} lists the uncertainties from the
Run~II lepton+jets measurements for CDF and {\dzero} that are based on  
the matrix-element technique~\cite{mass-lepton+jets-runII-cdf,mass-lepton+jets-runII-dzero}. These two
measurements provide most of the sensitivity to the combined $m_t$ 
result and are discussed below. 
Before explaining how each individual systematic uncertainty is estimated, 
we will first discuss how the uncertainties from different sources are propagated
to $m_t$ and how they are calculated using ensembles of pseudoexperiments.


%
\indent Uncertainties related to the performance of the detector 
and calibration of the reconstructed objects,  
such as JES, the modeling of jets, leptons, and triggers,
and calibration of the $b$-tagging algorithms, are evaluated by shifting 
the central values of their respective parameters by $\pm 1$ 
standard deviations ($\sigma$) that correspond to the uncertainties on each 
value. This is done using MC
$t\bar{t}$ events for $m_t = 172.5$~GeV. The integrations over the matrix 
element are performed again for each shifted sample and define shifts in $m_t $ 
that correspond to each independent source of systematic uncertainty. These uncertainties 
are not determined at other $m_t$ values,  
and it is assumed that their dependence on $m_t $ is minimal. 

\indent For uncertainties that arise from ambiguities in 
the modeling of the \ttbar\ signal, which include the uncertainties from initial-
and final-state radiation, higher-order QCD corrections, $b$-jet
hadronization, light-jet hadronization, the underlying-event model,
and color reconnection, we generate simulated \ttbar\ events using
alternative models also at $m_t = 172.5$~GeV.  
These events are processed through detector simulation
and are reconstructed, and the probability density is calculated by integration over the matrix elements.

For the uncertainties from the choice of parton distribution functions,
the ratio of contribution from quark annihilation and gluon fusion, and models 
for overlapping interactions, we reweight the
fully reconstructed simulated \ttbar\ MC events at 
$m_t =165, 170, 172.5, 175,$ and 180~GeV to reflect the uncertainty on 
the $\pm 1 \sigma$ range on each parameter and extract its impact on $m_t$. 

\indent Each method used to measure $m_t$ is calibrated
using $t\bar{t}$ MC events generated at $m_t = 165,
170, 172.5, 175, 180$~GeV, which provide the relationship between input 
and ``measured" masses. A straight
line is fitted to these values, representing a response
function that is used to correct the $m_t$ measurement in data. 

\indent Systematic uncertainties are evaluated using studies of ensembles 
of pseudoexperiments. 
For each of the shifted or reweighted sets of events, and those based on 
alternative models or different generated $m_t$, 
we create an ensemble of at least
1000 pseudoexperiments, by means of binomially smeared signal and
background fractions that match the expectation in the data sample and
with the total number of events in each pseudoexperiment equal to the number of
events observed in data. We use the ensembles of such pseudoexperiments 
to assess the difference between generated and measured mass
and to calibrate the
method of mass extraction. 

For the uncertainty on background, we change the fraction of background
events in the pseudoexperiments within their uncertainties and
remeasure the top-quark mass.


\begingroup
\squeezetable
\begin{table*}[!h!tb]
\begin{minipage}{4.5in}
\caption[ljets-uncertainties]{Individual components of uncertainty on
CDF and {\dzero} $m_t$ measurements in the lepton+jets channel for Run~II
data~\cite{mass-lepton+jets-runII-cdf,mass-lepton+jets-runII-dzero}.}
\label{ljets-uncertainties}
\vspace{0.05in}
\begin{ruledtabular}
\begin{tabular}{lcc}
           & \multicolumn{2}{c}{Uncertainty [GeV]} \vspace{0.05in}\\
Systematic & CDF~~~(5.6~fb$^{-1}$) & {\dzero}~~~(3.6~fb$^{-1}$)   \\
Source     &   $m_t = 173.00$~GeV  &   $m_t = 174.94$~GeV         \\
[1.0ex] \hline \\ [-1.5ex]
DETECTOR RESPONSE                     			  &        &        \\
~~~Jet energy scale 					  &        &        \\
~~~~~~Light-jet response (1)                             &  0.41  &  \nap  \\
~~~~~~Light-jet response (2)                             &  0.01  &  0.63  \\
~~~~~~Out-of-cone correction                              &  0.27  &  \nap  \\
~~~~~~Model for $b$ jets                                  &  0.23  &  0.07  \\
~~~~~~~~~{\it Semileptonic b decay}                       &   \bu  &   \bv  \\
~~~~~~~~~{\it b-jet hadronization}                        &   \bm  &   \bn  \\
~~~~~~Response to $b/q/g$ jets                              &  0.13  &  0.26  \\
~~~~~~{\em In situ} light-jet calibration                &  0.58  &  0.46  \\
[1.0ex]
~~~Jet modeling                                           &  0.00  &  0.36  \\
~~~~~~{\it Jet energy resolution}                         &   \z   &   \jk  \\
~~~~~~{\it Jet identification}                            &   \z   &   \jj  \\
[1.0ex]
~~~Lepton modeling                                        &  0.14  &  0.18  \\
[1.5ex]
MODELING SIGNAL			                          &        &        \\
~~~Signal modeling                                        &  0.56  &  0.77  \\
~~~~~~{\it Parton distribution functions}                 &   \pd  &   \pe  \\
~~~~~~{\it Quark annihilation fraction}                   &   \gf  &  \nap  \\
~~~~~~{\it Initial and final-state radiation}             &   \is  &   \iu  \\
~~~~~~{\it Higher-order QCD corrections}                  &  \nap  &   \ho  \\
~~~~~~{\it Jet hadronization and underlying event}        &   \jh  &   \ji  \\
~~~~~~{\it Color reconnection}                            &   \cs  &   \ct  \\
[1.0ex]
~~~Multiple interactions model                            &  0.10  &  0.05  \\
[1.5ex]
MODELING BACKGROUND   		                       &        &        \\
~~~Background from theory                              &  0.27  &  0.19  \\
~~~~~~{\it Higher-order correction for heavy flavor}   &   \hj  &   \hk  \\
~~~~~~{\it Factorization scale for W+jets}             &   \qs  &   \qt  \\
~~~~~~{\it Normalization to predicted cross sections}  &   \ti  &   \tj  \\
~~~~~~{\it Distribution for background}                &   \bs  &   \bt  \\
[0.5ex]
~~~Background based on data                            &  0.06  &  0.23  \\
~~~~~~{\it Normalization to data}                      &   \z   &   \tr  \\
~~~~~~{\it Trigger modeling}                           &   \z   &   \tr  \\
~~~~~~{\it $b$-tagging modeling}                       &   \z   &   \bb  \\
~~~~~~{\it Signal fraction for calibration}            &  \nap  &   \sg  \\
~~~~~~{\it Impact of multijet background on the calibration} &  \nap  &   \mc  \\
[1.5ex]
METHOD OF MASS EXTRACTION           		       &        &        \\
~~~Calibration method                               &  0.10  &  0.16  \\
[1.5ex] \hline \\ [-1.5ex]
STATISTICAL UNCERTAINTY	                               &  0.65  &  0.83  \\
[0.5ex]
UNCERTAINTY ON JET ENERGY SCALE                        &  0.80  &  0.83  \\
[0.5ex]
OTHER SYSTEMATIC UNCERTAINTIES                         &  0.67  &  0.94  \\
[1.5ex] \hline \\ [-1.5ex]
TOTAL UNCERTAINTY                                      &  1.23  &  1.50 
\vspace{0.02in}
\end{tabular}
\end{ruledtabular}
\end{minipage}
\end{table*}
\endgroup


%
\indent 
For the BLUE combination method, the uncertainties must be defined symmetrically 
around the central mass value, and this requirement determines part of the
following definitions of uncertainty.

For the uncertainties obtained in ensemble studies with shifted or reweighted
parameters, $m_t^+$ corresponds to the $+1\sigma$ shift in the input
parameter and $m_t^-$ corresponds to the $-1\sigma$ shift. The
systematic uncertainty on the value of $m_t$ from these parameters is
defined as {\mbox{$\pm \: |m_t^+ - m_t^-|/2$}}, unless both 
shifts are in the same direction relative to the nominal value,
in which case the systematic uncertainty is defined as the larger of 
$|m_t^+ - m_t|$ or $|m_t^- - m_t|$.

For the values obtained from a comparison between two or more 
models, the systematic uncertainty is taken as $\pm$ of the
largest difference among the resulting masses (without dividing by
two).


\vspace{0.2in}

\noindent {\bf 1.~~Jet energy scale}\\*[0.05in]
\indent The following seven terms (1.1 - 1.7) refer to the jet energy scale

\vspace{0.15in}
\noindent {\bf{1.1~~Light-jet response (1)}}\\*[0.05in]
\indent This uncertainty includes the absolute
calibration of the CDF JES for Run~I and Run~II and the smaller
effects on JES from overlapping 
interactions and the model for the underlying event.

CDF's calibration of the absolute jet energy scale uses the single-pion
response to calibrate jets in data and to tune the model of the calorimeter 
in the simulation. Uncertainties of these processes form the greatest part of the 
JES uncertainty. Small constant terms are added to account
for the model of jet fragmentation and for calorimeter simulation of
electromagnetically decaying particles, and to take into account small 
variations of the absolute calorimeter response over time.
The total resulting uncertainty on the absolute JES 
is 1.8\% for 20~GeV jets rising to 2.5\% for 150~GeV jets.

At high Tevatron instantaneous luminosities, more than one $p\bar{p}$ interaction
occurs during the same bunch crossing, and the average number
of interactions depends linearly on 
instantaneous luminosity and is changed from $\approx$ 1 to 8 between the
start and the end of Run~II. If the final-state particles 
from these extra $p\bar{p}$ interactions overlap with the
jets from a $t\bar{t}$ event, the energy of these jets is
increased, thereby requiring the correction.
The uncertainty
on this correction depends on vertex-reconstruction efficiency and
the rate for misidentifying vertices. The impact of these effects is checked on
data samples, including $W \rar e\nu$, minimum bias, and multijet events 
with a trigger threshold of 100~GeV. CDF finds an uncertainty of
0.05~GeV per jet. This uncertainty was estimated early in Run~II. With
increasing instantaneous luminosity, 
this correction was insufficient, and another  systematic
uncertainty term was introduced through the
``multiple-interactions-model'' term, which is described later.

CDF includes the impact of the underlying event on JES 
in this component of uncertainty. The proton and antiproton remnants
of the collision deposit energy in the calorimeter, and these
can contribute to the energy of the jets from \ttbar\
decay, which must be subtracted before $m_t$ 
can be measured accurately. CDF compares the ``Tune~A'' underlying-event
model~\cite{tuneA-cdf} in {\sc pythia}~\cite{pythia2} with the {\sc
jimmy} model~\cite{jimmy,tuneA-jimmy} in {\sc herwig}~\cite{herwig2}
using isolated tracks with $p_T > 0.5$~GeV. The data agree well with
Tune~A, which is expected since it was tuned to CDF data, but differ
from {\sc jimmy} by about 30\%. This difference is propagated to the
absolute calibration of JES and yields a 2\%
uncertainty for low-$p_T$ jets and less than 0.5\% for 35~GeV jets.

MC \ttbar\ events are generated by CDF with jet energies shifted 
by the above three uncertainties, and the resulting shifts in $m_t$ 
are used to estimate the uncertainty. The overall
uncertainty on $m_t$ from these
combined sources is 0.24\% for lepton+jets, 0.22\% for alljets, 1.18\%
for CDF Run~II dilepton data, and 0.26\% for {\met}+jets for Run II data of CDF.


\vspace{0.15in}
\noindent {\bf{1.2~~Light-jet response (2)}}\\*[0.05in]
\indent This uncertainty term represents almost all
parts of {\dzero} Run~I and Run~II 
calibrations of JES. 
The absolute energy scale for jets in data is calibrated
using $\gamma$+jet data with photon $p_T > 7$~GeV and $|\eta_{\gamma}| < 1.0$,
and jet $p_T > 15$~GeV and $|\eta_{jet}| < 0.4$, using the ``{\met} projection
fraction'' method~\cite{jet-energy-scale-dzero}. Simulated samples of $\gamma$+jets and $Z$+jets
events are compared to data and used to correct the energy scale for jets
in MC events. The JES is also corrected as a function
of $\eta$ for forward jets relative to the central jets 
using $\gamma$+jets and dijets data. Out-of-cone
particle scattering corrections are determined with $\gamma$+jets data and simulated
events, without using overlays of underlying events, to avoid double counting of 
this effect. Templates of deposited energy are formed for particles belonging to and not
belonging to a jet using 23 annular rings around the jet axis for ${\cal
R}(y,\phi) = \sqrt {(\Delta y)^2 + (\Delta\phi)^2}  \le 3.5$. All of these
calibration steps are combined, and the total 
uncertainty on JES is calculated for light jets and heavy-flavor jets (independent 
of the type of jet). The resulting {\dzero} 
uncertainty on $m_t$  for Run~II lepton+jets events is 0.36\% and 
0.86\% for dilepton data.

This uncertainty term also includes the relative jet energy
correction as a function of jet $\eta$ for CDF. This is measured
using dijet data, along with {\sc pythia} and {\sc herwig} simulations of
\ttbar\ events generated with shifted jet energies, and lead to the
following uncertainties on Run~II measurements of $m_t$:
0.01\% for lepton+jets, 0.02\% for alljets, 0.34\% for dileptons, and
0.03\% for {\met}+jets.


\vspace{0.15in}
\noindent {\bf{1.3~~Out-of-cone corrections}}\\*[0.05in]
\indent For all CDF measurements and for {\dzero} Run~I, this
uncertainty component accounts for energy lost outside the jet
reconstruction cone and uses the difference between two models of
light-quark and gluon fragmentation and simulation of the underlying event. 
{\dzero} changed the way it measures the out-of-cone
uncertainty between Run~I and Run~II, and this uncertainty for
{\dzero} Run~II measurements is therefore included in the
light-jet response (2) term, described previously.

Energy is lost from the cone of jet reconstruction when a quark or gluon
is radiated at a large angle relative to the original parton direction, or when the
fragmentation shower is wider than the cone, or when
low momentum particles are bent out of the cone by
the axial  magnetic field of the detector. Energy is gained in the cone from
initial-state radiation and from remnants of spectator partons,
called collectively at CDF the underlying event. The two models
compared by CDF in Run~II are {\sc pythia} with Tune~A for the
underlying event and {\sc herwig} with the {\sc jimmy} modeling of the
underlying event. For the narrow cone size of ${\cal R} = 0.4$ used in
measurements of $m_t$, more energy is lost from the cone
than gained. The correction is measured using {\sc pythia} dijet
events and data in the region $0.4 < {\cal R} \le 1.3$.
A small constant is added to compensate for energy outside the ${\cal R} > 1.3$  region (``splash out'').
 The correction is largest for jets at low
transverse momentum: +18\% for $p_T = 20$~GeV jets and $< 4\%$
for jets with $p_T > 70$~GeV. A detailed description of the method can
be found in Ref.~\cite{jet-energy-scale-cdf}.

The uncertainty on these corrections is measured by comparing $\gamma+{\rm jets}$ data
to the two simulations. The largest difference
between either of the models and data is taken as the uncertainty
(the difference between the two models is very small). For
jets with $p_T = 20$~GeV, the uncertainty on the jet energy scale is
6\%, and for jets above 70~GeV, it is 1.5\%. These translate into
uncertainties on CDF Run~II $m_t$ measurements of 0.16\%
for the lepton+jets measurement, 0.14\% for alljets, 1.25\% for
dileptons, and 0.12\% for {\met}+jets.

 
\vspace{0.15in}
\noindent {\bf{1.4~~Energy offset}}\\*[0.05in]
\indent This uncertainty term is specific to {\dzero}  Run~I
measurements. It includes the uncertainty arising from uranium decays noise
in the calorimeter and from the correction for multiple interaction to JES. 
These lead to uncertainties in $m_t$ of
0.72\% for lepton+jets and 0.77\% for dilepton events. In Run~II, the
integration time for the calorimeter electronics is short, after the 
upgrade to shorter bunch-crossing time (3.5~$\mu$s to
396~ns). This effect results in a negligible uncertainty on the offset for
{\dzero} Run~II measurements of $m_t$.


\vspace{0.15in}
\noindent {\bf{1.5~~Model for $\mathbi{b}$ jets}}\\*[0.05in]
\noindent {(i)~{\it Semileptonic b decay}}\\*[0.05in]
\indent The uncertainty on the semileptonic branching fraction 
$(10.69 \pm 0.22) \times 10^{-2}$ (PDG 2007 values)
of $B$~hadrons affects the value of $m_t$. Both collaborations
reweight \ttbar\ events by $\pm$ the uncertainty on the
central value ($\pm 2.1$\%), and take half the resulting mass difference as
the uncertainty on $m_t$: 0.09\% for CDF and 0.03\% for {\dzero}.

\vspace{0.1in}
\noindent {(ii)~{\it b-jet hadronization}}\\*[0.05in]
\indent For its nominal $m_t$ measurements, CDF uses the
default {\sc pythia} model of $b$-jet fragmentation
based on the Bowler model~\cite{bowler-b-decay}
($r_q = 1.0$, $a = 0.3$, $b = 0.58$), where $r_q$ is the Bowler fragmentation-function
parameter and $a$ and $b$ are Lund fragmentation-function parameters.
{\dzero} uses a model with these parameters tuned to
data from ALEPH, DELPHI, and OPAL~\cite{b-tuning}
($r_q = 0.897\pm0.013$, $a = 1.03\pm0.08$, $b = 1.31\pm0.08$).
To measure the uncertainty on these models, CDF compares its $m_t$ 
values to those measured with the LEP parameters used by {\dzero}, 
and to those from the SLD experiment at SLC~\cite{b-tuning} ($r_q =
0.980\pm0.010$, $a = 1.30\pm0.09$, $b = 1.58\pm0.09$). {\dzero}
compares the measured $m_t$ with the LEP parameters to the one from
SLC. The resulting uncertainties on the $m_t$ extracted from the
lepton+jets channel are 0.09\% for CDF and 0.03\% for {\dzero}.

For  some analyses, the determination of the uncertainties in (i) and (ii) may be affected 
 by statistical fluctuations of the MC samples.


\vspace{0.15in}
\noindent {\bf{1.6~~Response to $\mathbi{b/q/g}$ jets}}\\*[0.05in]
\indent The calibrations of JES described in the first two
paragraphs of the Appendix are derived on samples dominated by ``light-quark'' 
and gluon jets and applied to all jets. However, the calorimeter response
to heavy-flavor jets differs in that these
particles often decay semileptonically, and the $b$ jet will have some
energy lost through the escaping neutrino. Bottom quark jets can also contain an electron
that showers in a pattern different than for hadronic particles, or the jet may contain a muon
that neither produces a shower nor gets absorbed in the calorimeter.
Bottom  jets also differ from light jets in the distribution of
their shower and particle content. Since every $t\bar{t}$ event
contains two $b$~jets, it is important to understand their energy
calibration after the application of the previous overall corrections.

CDF measures an uncertainty from the difference between the $b$-jets
response and light-flavor jets response in Run~II. CDF
takes sets of MC $t\bar{t}$ events and cluster particles into
jets classifying each such particle jet as a $b$~jet or a light
jet~\cite{jet-algorithm-cdf}. Single-particle response
for data and for MC events are applied to the formed particle
jets to predict the energy measured in the calorimeter. A double ratio
is calculated: $(p_T^{\rm data}/p_T^{\rm MC})_{b{\rm ~jets}} /
(p_T^{\rm data}/p_T^{\rm MC})_{\rm light~jets}$, which is found to be 1.010. The
uncertainty on $m_t$ is measured by generating new \ttbar\
samples with the $b$-jet scale shifted by this $1\%$ difference, which
results in 0.1\% uncertainty in $m_t$ for the lepton+jets measurement.

For Run~II measurements, {\dzero} corrects the transverse-momentum
distributions of jets differently in four regions of detector pseudorapidity to
make the MC response match that in data (after the main JES
calibration) as a function of jet flavor: $b$~jets,
light-quark jets ($u$, $d$, $s$, $c$), and
gluon~jets~\cite{mass-lepton+jets-runII-dzero}. The correction
functions are shifted up and down by their uncertainties, and the extracted shifts in $m_t$
are used to define the resulting uncertainty on $m_t$ of
0.15\% for the lepton+jets measurement and 0.23\% for
the dilepton measurement.


\vspace{0.15in}
\noindent {\bf 1.7~~{{\em In situ} light-jet calibration}}\\*
[0.05in]
\indent In $t\bar{t}$ events where one or both $W$~bosons decay
to $q \bar{q}'$, the world average value of $M_W$~\cite{particle-data-book}
is used to constrain the jet
energy scale for light-quark jets {\em in situ}~\cite{in-situ-cdf,in-situ-dzero}. 
CDF and {\dzero} perform
simultaneous measurements of $m_t$ and $M_W$, and
fit a linear function to the JES for light-quark jets that is
applied to all the jets to improve precision of $m_t$.

CDF measures the {\em in situ} rescaling factor independently in their
lepton+jets, alljets, and {\met}+jets analyses, and so these terms are
uncorrelated. {\dzero} applies the rescaling derived from their
lepton+jets measurement to dilepton events, and these
uncertainties are therefore correlated.

The uncertainty from the {\em in situ} calibration is determined through
a two-dimensional minimization of a likelihood that is a function of top-quark mass  and JES.
The extracted JES is then shifted relatively to its
measured central value, and a one-dimensional fit is performed to the top-quark mass. 
The difference in quadrature between the uncertainty on $m_t$ from the first and second fits is 
taken as the uncertainty on $m_t$ from the
{\em in situ} calibration, giving 0.34\% for CDF's lepton+jets
measurement, 0.27\% for {\dzero}'s lepton+jets result, 0.55\% for
CDF's alljets, 0.89\% for their {\met}+jets measurement, and 0.32\%
for {\dzero}'s dilepton measurement.


\vspace{0.15in}
\noindent {\bf{2.~~Jet modeling}}\\*[0.05in]
\indent Applying jet algorithms to MC events, CDF finds that 
the resulting efficiencies and resolutions closely match those in data.
The small differences propagated to $m_t$ lead  to a negligible uncertainty of
$0.005$~GeV, which  is then ignored. {\dzero} proceeds as follows.

\vspace{0.1in}
\noindent {(i)~{\it Jet energy resolution}}\\*[0.05in]
\indent The modeling of the jet energy resolution is corrected in
{\dzero} to match that in data. The value of $m_t$ is then
remeasured using MC samples with jet energy resolution corrections
shifted up and down by their uncertainties, resulting in an uncertainty
on $m_t$ of 0.18\%.

\vspace{0.1in}
\noindent {(ii)~{\it Jet identification}}\\*[0.05in]
\indent {\dzero} applies correction functions to MC events
to match the jet identification efficiency in
data. The uncertainty on $m_t$ is estimated by reducing
the corrections by $1 \sigma$ and remeasuring the mass in
the adjusted MC samples. The efficiency can only be shifted
down and not up because jets can be removed from the simulated events but
not added. The uncertainty on $m_t$ is therefore set to $\pm$  the single-sided
shift and is 0.15\%.


\vspace{0.15in}
\noindent {\bf{3.~~Lepton modeling}}\\*[0.05in]
\noindent {(i)~{\it Momentum scale for leptons}}\\*[0.05in]
\indent In Run~II, the electron and muon channels for CDF and the muon
channels for {\dzero} are used to calibrate the lepton momentum scales by
comparing the invariant dilepton mass $m_{\ell1\ell2} =
\sqrt{(E_{\ell1}+E_{\ell2})^2 - (p_{\ell1} + p_{\ell2})^2}$ for $J/\psi
\rar \ell\ell$ and $Z \rar \ell\ell$ decays in MC events with data. The
positions of the resonances observed in the $m_{\ell\ell}$ distributions reflect the 
absolute momentum scales for the leptons.
CDF and {\dzero} perform a linear fit as a function of the mean value of transverse momentum to
the two mass points (3.0969~GeV and 91.1876~GeV~\cite{particle-data-book}), assuming that any
mismatch is attributable to an uncertainty in the calibration of the magnetic field. {\dzero} also fits a
quadratic relation, assuming that the difference in scale arises from
misalignment of the detector. The value of $m_t$ is measured using
MC \ttbar\ ensembles without rescaling lepton $p_T$ and with lepton $p_T$ values rescaled 
using these fitted relations. Half of the 
largest difference in extracting $m_t$ is taken as its systematic uncertainty resulting from
the lepton $p_T$ scale. For muon measurements from {\dzero},
the largest shift is observed for the linear parametrization. In Run~I,
this source of uncertainty was neglected as it was negligible relative to other sources
of uncertainty.

In {\dzero} Run~II measurement of the $W$-boson mass in the electron decay
channel, it was found that 0.26 radiation length of material was left out in
the {\sc geant} modeling of the
solenoid~\cite{W-mass-dzero}. The $Z$-boson mass peak was used to
calculate a quadratic correction to the electron energy by comparing
MC events generated with additional solenoid material to
data. This correction was then propagated to the $m_t$
measurement.

The uncertainties on the $m_t$ measurements from the lepton
momentum scale are 0.08\% for CDF lepton+jets
measurements and 0.10\% for {\dzero}, and 0.16\% for CDF dilepton
measurements and 0.28\% for {\dzero} dilepton results.

\vspace{0.1in}
\noindent {(ii)~{\it  Lepton momentum resolution}}\\*[0.05in]
\indent The muon momenta in simulated events at {\dzero} are
smeared to match the resolution in data. The
uncertainty on this correction corresponds to an
uncertainty on $m_t$ of 0.17\%.


\vspace{0.2in}
\noindent {\bf 4.~~Signal modeling}\\*[0.05in]
\noindent {(i)~{\it Parton distribution functions}}\\*[0.05in]
\indent In Run~I, the uncertainties from choice of PDF 
are determined by measuring the change in $m_t$
using the MRSA$^{\prime}$ set~\cite{mrsap} instead of
MRSD$_0^{\prime}$~\cite{mrsd0p} or CTEQ4M~\cite{cteq4m}, and are
found to be negligible.

In Run~II, the uncertainty is measured by CDF by comparing CTEQ5L
results with MRST98L~\cite{mrst98}, by changing the value of
$\alpha_s$ in the MRST98L model, and by varying the 20 eigenvectors in
CTEQ6M~\cite{cteq6}. The total uncertainty is obtained by combining
these sources in quadrature. {\dzero} measures this uncertainty by
reweighting the {\sc pythia} model to match possible excursions in the
parameters represented by the 20 CTEQ6M uncertainties
and taking the quadratic sum of the
differences. The resulting uncertainty on $m_t$ is 0.08\%
for CDF and 0.14\% for {\dzero}.

\vspace{0.1in}
\noindent{(ii)~{\it Fractional contributions from quark annihilation and gluon fusion}}\\*
[0.05in]
\indent In Run~I, this source of uncertainty in \ttbar\ production is not considered. In
Run~II, CDF estimates the effect on $m_t$ by reweighting the gluon fusion
fraction in  the {\sc pythia} model from 5\% to 20\%~\cite{gluon-fusion-cdf}. 
The uncertainty on $m_t$ is found to be 0.02\%. This
uncertainty is included by {\dzero} in the systematic component (iv)
below, where the effects of higher-order QCD corrections are discussed.

\vspace{0.1in}
\noindent {(iii)~{\it Initial- and final-state radiation}}\\*
[0.05in]
\indent Initial- and final-state radiation refers to additional gluons
radiated from the incoming or outgoing partons or from the top quarks.
Jets initiated by these gluons affect the measured value of $m_t$
because they can be misidentified as jets from the final-state
partons in top-quark decay.
Extensive checks were performed in Run~I measurements to assess the
effects of initial- and final-state radiation by
varying parameters in {\sc herwig}.

In Run~II, uncertainties from initial- and final-state radiation
are assessed by both collaborations using a CDF
measurement~\cite{isr-fsr-cdf} in Drell-Yan dilepton events that have the
same $q\bar{q}$ initial state as most $t\bar{t}$ events, but no
final-state radiation. The mean $p_T$ of the produced dilepton
pairs is measured as a function of the
dilepton invariant mass, and the values of $\Lambda_{\rm
QCD}$ and the $Q^2$ scale in the MC that
matches best the data when extrapolated to the $t\bar{t}$ mass region are found. CDF's
best-fit values are $\Lambda_{\rm QCD}$ (5 flavors) = 292~MeV with
$0.5 \times Q^2$ and $\Lambda_{\rm QCD}$ (5 flavors) = 73~MeV
with $2.0 \times Q^2$ for $\pm \sigma$ excursions around the mean dilepton $p_T$ values. 
Since the initial and final-state
shower algorithms are controlled by the same QCD evolution
equation~\cite{dglap-evolution}, the same variations of $\Lambda_{\rm
QCD}$ and $Q^2$ scale are used to estimate the effect of final-state
radiation. The resulting uncertainty for modeling of the initial-
and final-state radiation is 0.09\% for CDF and 0.15\% for
{\dzero}. The correction algorithm does not distinguish
between ``soft'' (out-of-cone) and ``hard'' (separate jet) radiation,
and there is therefore some
overlap between the uncertainty on $m_t$ for the out-of-cone jet
energy correction and for gluon radiation. There is also some overlap
between the uncertainty for initial- and final-state radiation and
the uncertainty on higher-order QCD corrections for high-$p_T$
radiation.

\vspace{0.1in}
\noindent {(iv)~{\it Higher-order QCD corrections}}\\*[0.05in]
\noindent 
Higher-order QCD corrections to \ttbar\ production are not used 
for Run~I measurements, as only LO generators were
available at that time. {\dzero} measures higher-order jet-modeling uncertainties in
Run~II by comparing $m_t$ extracted with {\sc alpgen} and {\sc herwig} for evolution
and fragmentation to the value obtained from events generated with {\sc
mc@nlo}~\cite{mc@nlo}, which uses {\sc herwig} parton showering with a
NLO model for the hard-scattering process. This component of uncertainty also includes
(for {\dzero}) the uncertainty from the fraction of quark-antiquark to
gluon-gluon contributions to the initial state. CDF also studies differences
in $m_t$ using {\sc mc@nlo} and finds that the uncertainties in
distributions in the number of jets and the transverse momentum of the
$t\bar{t}$ system overlap with the uncertainty from
initial- and final-state radiation. Future measurements of $m_t$
are expected to treat these uncertainties separately. The uncertainty on $m_t$
from higher-order contributions and initial-state $q \bar{q}/gg$ ratio
is 0.14\% for {\dzero}.

\vspace{0.1in}
\noindent {(v)~{\it Jet hadronization and underlying event}}\\*
[0.05in]
\indent In Run~I, CDF measured the uncertainty in the model for parton showering
and hadronization and the underlying-event by comparing the value of $m_t$
based on {\sc herwig} to that on {\sc
pythia}~\cite{pythia1}, and {\dzero} compared {\sc herwig} results to
those from {\sc isajet}~\cite{isajet}.

In Run~II, CDF estimates these uncertainties by comparing $m_t$ obtained
using {\sc pythia} with Tune~A of the underlying-event
model to results from {\sc herwig} with a tuned implementation of the
underlying-event generator {\sc jimmy}. {\dzero} estimates these
uncertainties by comparing identical sets of hard-scatter events from
{\sc alpgen} coupled to {\sc herwig} instead of to {\sc pythia}. For
the uncertainty on $m_t$, this corresponds to 0.40\% for CDF and
0.33\% for {\dzero}.

\vspace{0.1in}
\noindent {(vi)~{\it Color reconnection}}\\*[0.05in]
\indent There are up to six final-state quarks in $t\bar{t}$
events, in addition to initial and final-state radiation. When
hadronization and fragmentation occur, there are color interactions among
these partons and the color-remnants of the proton and antiproton.
This process is referred to as ``color reconnection''. It changes the
directions and distributions of final-state jets~\cite{color-reconnection-webber,color-flow-dzero},
which affects the reconstructed value of $m_t$~\cite{color-reconnection-skands}. 

The uncertainty on color reconnection was not evaluated for Run~I
because appropriate MC tools were not available at that
time. Both collaborations estimate this effect in Run~II by comparing
the value of $m_t$ extracted from ensembles of \ttbar\ events generated by {\sc
pythia} using the difference between two parton shower simulations: (i) angular ordering for jet showers 
(same as used in the nominal $m_t$ measurements) using the {\sc {\mbox{A-pro}}}
underlying-event model (Tune~A but updated using the ``Professor'' tuning
tool~\cite{professor}), and (ii) {\sc {\mbox{ACR-pro}}}. {\sc {\mbox{ACR-pro}}} is identical to {\sc
{\mbox{A-pro}}} except that it includes color reconnection in the
model. The resulting uncertainties on $m_t$ are 0.32\% for CDF and 0.16\% for
{\dzero}.


\vspace{0.15in}
\noindent {\bf{5.~~Multiple interactions model}}\\*[0.05in]
\indent Monte Carlo simulated events are overlaid with Poisson-distributed
low-$p_T$ events ({\sc pythia} MC events for CDF,
``zero-bias" data for {\dzero}) to simulate the presence of simultaneous additional
$p\bar{p}$ interactions. The mean number of overlaid events is chosen
at the time of event generation, but in data, the number of
such interactions changes with instantaneous
luminosity of the Tevatron.

CDF measures $m_t$ as a function of the number of
multiple interactions, finding a change of $0.07\pm0.10$~GeV per
primary vertex. For CDF's measurements, the average number of primary
vertices in data is 2.20 and for simulated events it is 1.85, leading
to an uncertainty on $m_t$ of 0.02\%. CDF adds to this in
quadrature a term to cover the difference in jet energy response as a
function of the number of multiple interactions of 0.06\%, giving a
total uncertainty of 0.06\%.

{\dzero} reweights the simulated events to make the instantaneous
luminosity distribution match that in data. The resulting
uncertainty on $m_t$ is 0.03\%.


\vspace{0.2in}
\noindent {\bf{6.~~Background from theory}}\\*[0.05in]
\noindent {(i)~{\it Higher-order correction for heavy flavor}}\\*
[0.05in]
{\dzero} corrects the leading-log $W+$jets cross section from {\sc alpgen}
to NLO precision before normalizing this background to data.
This increases the fraction of $Wb\bar{b}$ and $Wc\bar{c}$ events
in $W$+jets by a factor of $1.47 \pm 0.50$. CDF
normalizes the $W$+heavy-flavor jets background to data independent
of the other components in $W$+jets, which has a similar effect. The
resulting uncertainties on $m_t$ are 0.11\% for CDF and
0.04\% for {\dzero}.

\vspace{0.1in}
\noindent {(ii)~{\it Factorization scale for $W$+jets}}\\*
[0.05in]
\indent The transverse momenta of the jets in $W$+jets events are
sensitive to the factorization and renormalization scales chosen for
the calculations. These two scales are set equal to each other, with
$Q^2 = M_W^2 + \sum p_T^2$. To determine the uncertainty on $m_t$,
the scale is changed from $(Q/2)^2$ to $(2 \times Q)^2$, the MC
events regenerated, and the mass remeasured. Changing the scale
does not affect the fraction of $W$+jets in the model but
does affect the transverse-momentum distributions of the jets.
The uncertainties on $m_t$ are 0.02\% for CDF and 0.09\% for
{\dzero}.

\vspace{0.1in}
\noindent {(iii)~{\it Normalization to predicted cross sections}}\\*
[0.05in]
\indent CDF divides the background into seven independent parts: $W$+heavy-flavor 
jets, $W$ + light-flavor jets, single-top $tqb$ and $tb$,
$Z$+jets, dibosons ($WW$, $WZ$, and $ZZ$), and multijet contributions. This
uncertainty term covers the normalization of the components modeled
with MC simulated events (not multijets). The small backgrounds from
single-top, $Z$+jets, and diboson production are normalized to
NLO calculations. The uncertainties on the
cross sections are 10\% for $tqb$, 12\% for $tb$, 14\% for $Z$+jets,
and 10\% for dibosons. The $W$+jets background is normalized to data
before implementation of $b$ tagging, using a fit to the distribution for \met\
in the event. The uncertainty on this normalization cannot easily be  disentangled from the 
other sources, and so it is kept in this category. The combined uncertainty on $m_t$ from
these normalizations is 0.09\%.

{\dzero} also normalizes single-top, $Z$+jets, and diboson contributions, in
all analysis channels, and Drell-Yan in the dilepton channel, to
next-to-leading-order cross sections, using values from the
{\sc mcfm} event generator~\cite{mcfm}. The uncertainties on the cross
sections take into account the uncertainty on the PDF
and on the choice of factorization and renormalization scales,
which together propagate through to $m_t$
an uncertainty of 0.04\%.

\vspace{0.1in}
\noindent {(iv)~{\it Background differential distributions}}\\*
[0.05in]
\indent  For  CDF,  different methods were used to estimate the  uncertainty  attributable to  the 
overall background shape. In the  recent  lepton+jets  analysis, 
this uncertainty was assessed by dividing randomly the background events 
into subsets, building the background likelihood  from one of the subsets, 
and reconstructing the $m_t$ from the second subset. 
In the next step, the difference in $m_t$ obtained from the second
subset and the nominal  $m_t$  value is evaluated. 
This contributes an uncertainty of 0.03\%. CDF also estimates
an uncertainty from the limited MC statistics used to measure
the background. This yields an additional 0.03\% uncertainty on $m_t$.

For {\dzero}, the $p_T$ and $\eta$ distributions
of jets in $W$+jets events do not fully reproduce those in
data. An uncertainty to cover these deviations is based on the
difference between the model for background and data in 
the $\eta$ distribution of the third jet in three-jet events. The resultant uncertainty 
on $m_t$ is 0.09\%.


\vspace{0.15in}
\noindent {\bf{7.~~Background based on data}}\\*[0.05in]
\noindent {(i)~{\it Normalization to data}}\\*[0.05in]
\indent In the lepton+jets, alljets, {\met}+jets, and decay-length
channels, backgrounds from multijet events are normalized to data.
In the lepton+jets
analyses at {\dzero}, the $W$+jets background model is combined with the contribution from multijet
events, and both are normalized simultaneously to data, so that their
uncertainties in normalization are anticorrelated.
In dilepton analyses at CDF, the
Drell-Yan background is normalized to data. For the lepton+jets
analyses, CDF uncertainty on $m_t$ from the normalization of the multijet
backgrounds to data is 0.03\%, and {\dzero}'s uncertainty for the
normalization of $W$+jets and multijets to data is 0.13\%.

\vspace{0.1in}
\noindent {(ii)~{\it Trigger modeling}}\\*[0.05in]
\indent CDF expects a negligible uncertainty on $m_t$ from the modeling of the trigger.
{\dzero} simulates the trigger turn-on efficiencies for MC
events by applying weights as a function of the transverse
momentum of each object in the trigger. The uncertainty is measured by
setting all the trigger efficiencies to unity and recalculating the
value of $m_t$, which shifts $m_t$ by 0.03\%.

\vspace{0.1in}
\noindent {(iii)~{\it $b$-tagging modeling}}\\*[0.05in]
\indent CDF applies the $b$-tagging algorithm directly to
MC events and finds that any difference between the $b$-tagging behavior 
in MC and data has a negligible impact on the measurement of $m_t$. 
{\dzero} applies the $b$-tagging
algorithm directly to MC events for recent Run~II
measurements. Previously $b$-tagging was simulated with tag probability, 
and in Run~I, as {\dzero} did not have a silicon tracker, nonisolated muons 
were used to identify $b$ jets. The tagging efficiency for simulated events is made
to match that in data by randomly dropping $b$~tags for $b$ and
$c$ jets, while assigning a per jet weight for tagging light-flavor
jets as $b$ jets. The uncertainties for these corrections are determined by
shifting the efficiencies for tagging $b$ and $c$ jets by 5\% and by 20\% for
light jets, which introduces an uncertainty on $m_t$ of 0.06\%.

\vspace{0.1in}
\noindent {(iv)~{\it Signal fraction for calibration}}\\[0.05in]
\indent {\dzero} measures the impact of the uncertainty in the ratio
of signal to background events, which affects the calibration of $m_t$.
Changing the signal fraction within uncertainty results in
an uncertainty on $m_t$ of 0.06\%.

\vspace{0.1in}
\noindent {(v)~{\it Impact of multijet background on the calibration}}\\*
[0.05in]
\indent Multijet background events are not used in {\dzero}
samples that determine the calibration of $m_t$ for the lepton+jets
measurement since the background probability for such events is much larger than the
signal probability. The assumption that this has a
small effect on $m_t$ is tested by selecting a multijet-enriched sample of
events from data (by inverting the lepton isolation criteria) and adding
these events when deriving the calibration. Applying this alternative calibration 
to data indicates that $m_t$ can shift by an uncertainty of 0.08\%.


\vspace{0.2in}
\noindent {\bf{8.~~Calibration method}}\\*[0.05in]
\indent  Monte Carlo \ttbar\ ensembles are generated at different values of input $m_t$
($m_t = 165, 170, 172.5, 175, 180$~GeV), 
and calibrations relate the input masses for \ttbar\
events to the extracted masses using a straight line. 
For some of the $m_t$ measurements, there is an
additional {\em in situ} calibration of the JES to
the light quarks in $W$-boson decay, which is then applied to all jets. 
The uncertainties from both calibrations are
propagated to the uncertainty on $m_t$, which for CDF
are 0.04\% and 0.05\%, respectively, giving a total of
0.06\%. For {\dzero}, the uncertainty on $m_t$ is 0.13\%.


%
%
%



\begin{thebibliography}{999}
\bibitem{pseudorapidity}
The pseudorapidity $\eta$ is defined as: $\eta = -\ln [\tan (\theta/2)]$, where $\theta$
is the polar angle with respect to the proton beamline.
\end{thebibliography}
\end{document}